\documentclass[12pt]{article}
\usepackage{amssymb,amsmath,amsthm,amsxtra,amsfonts}
\usepackage{geometry}
\usepackage{aliascnt}
\usepackage{color}
\usepackage[usenames,dvipsnames,table]{xcolor}
\usepackage{subcaption}
\usepackage{bbm}
\usepackage{dsfont}
\usepackage{graphicx}
\usepackage{mathrsfs}
\usepackage{tikz}
\usetikzlibrary{plotmarks,arrows,calc,patterns}
\usepackage[round]{natbib}
\usepackage[colorlinks=true,citecolor=Blue,urlcolor=Blue,linkcolor=Blue]{hyperref}
\usepackage{algorithm,algpseudocode}
\usepackage{threeparttable}
\usepackage{booktabs}
\usepackage{rotating}
\usepackage{algorithm}
\usepackage{longtable}

\newtheorem*{assumption*}{Assumption}

\newcommand{\R}{\ensuremath{\mathbf{R}}}

\newcommand{\E}{\ensuremath{\mathbb{E}}}

\begin{document}
	\title{Panel Data Nowcasting: The Case of Price-Earnings Ratios\footnote{We benefited from comments by Rudy De Winne, Geert D'Haene, Max Farrell, Christian Hafner, Peter Reinhard Hansen, Dacheng Xiu, and participants at the 2021 SoFiE UC San Diego conference, 26th International Panel Data Conference, Data Science and Machine Learning workshop at the University of Amsterdam, the 2022 IAAE Conference, King's College, London, and the 2022 Vienna Copenhagen Conference on Financial Econometrics. This work was in part completed when Jonas Striaukas was a Research Fellow at Fonds de la Recherche Scientifique FNRS.}}
	
	\author{Andrii Babii\thanks{University of North Carolina at Chapel Hill - Gardner Hall, CB 3305 Chapel Hill, NC 27599-3305. Email: babii.andrii@gmail.com.} \and Ryan T. Ball\thanks{Stephen M. Ross School of Business, University of Michigan, 701 Tappan Street, Ann Arbor, MI 48109. Email: rtball@umich.edu.} \and Eric Ghysels\thanks{Department of Economics and Kenan-Flagler Business School, University of North Carolina--Chapel Hill. Email: eghysels@unc.edu.} \and Jonas Striaukas\thanks{Department of Finance, Copenhagen Business School, Frederiksberg, Denmark. Email: jonas.striaukas@gmail.com.}}
	
	\maketitle
	
	\begin{abstract}
		\noindent {\footnotesize The paper uses structured machine learning regressions for nowcasting with panel data consisting of series sampled at different frequencies. Motivated by the problem of predicting corporate earnings for a large cross-section of firms with macroeconomic, financial, and news time series sampled at different frequencies, we focus on the sparse-group LASSO regularization which can take advantage of the mixed frequency time series panel data structures. Our empirical results show the superior performance of our machine learning panel data regression models over analysts' predictions, forecast combinations, firm-specific time series regression models, and standard machine learning methods.}
	\end{abstract}
	
	\noindent%
	\vfill
\thispagestyle{empty}

\setcounter{page}{0}

\newpage

\section{Introduction}\label{sec:intro}
	
Nowcasting is intrinsically a mixed frequency data problem as the object of interest is a low-frequency data series --- observed say quarterly --- whereas real-time information --- daily, weekly or monthly --- during the quarter can be used to assess and potentially continuously update the state of the low-frequency series, or put differently, {\it nowcast} the series of interest. Traditional methods being used for nowcasting rely on dynamic factor models which treat the underlying low-frequency series of interest as a latent process with high-frequency data noisy observations. These models are naturally cast in a state-space form, and inference can be performed using standard techniques (in particular the Kalman filter, see \cite{banbura2013now} for a recent survey). 

\medskip

Things get more complicated when we are operating in a data-rich environment {\it and} we have many target variables. Put differently, we are no longer interested in  nowcasting a single key series such as the GDP growth where we could devote a lot of resources to that particular series. A good example is corporate earnings nowcasting for a large cross-section of corporate firms. The fundamental value of equity shares is determined by the discounted value of future payoffs. Every quarter investors get a glimpse of firms' potential payoffs with the release of corporate earnings reports. In a data-rich environment, stock analysts have many indicators regarding future earnings that are available much more frequently. \cite{ball2018automated} took a first stab at automating the process using MIDAS regressions. Since their original work, much progress has been made on machine learning (ML) regularized mixed frequency regression models. 

\medskip

In the context of earnings, we are potentially dealing with a large set of individual firms for which there are many predictors. From a practical point of view, this is clearly beyond the realm of nowcasting using state space models. In the current paper, we significantly expand the tools of nowcasting in a data-rich environment by exploiting panel data structures. Panel data regression models are well suited for the firm-level data analysis as both the time series and cross-sectional dimensions can be exploited. In such models, time-invariant firm-specific effects are typically used to capture cross-sectional heterogeneity in the data. This is combined with regularized regression machine learning methods which are becoming increasingly popular in economics and finance as a flexible way to model predictive relationships via variable selection.  We focus on the panel data regressions in a high-dimensional data setting where the number of covariates could be large and potentially exceed the available sample size. This may happen when the number of firm-specific characteristics, such as textual analysis news data or firm-level stock returns, is large, and/or the number of aggregates, such as market returns, macro data, etc., is large. 

\medskip

Our paper relates to several existing papers in the literature. \cite{khalaf2020dynamic} consider low-dimensional dynamic mixed frequency panel data models but do not deal with high-dimensional data situations in the context of nowcasting or forecasting. Similarly, \cite{fosten2019panel} consider nowcasting with a mixed-frequency VAR panel data model, but not in the context of a high-dimensional data-rich environment that we are interested in here. \cite{babii2022machine} introduce the sparse-group LASSO (sg-LASSO) regularization machine learning methods for heavy-tailed dependent panel data regressions potentially sampled at different time series frequencies. They derive oracle inequalities for the pooled and fixed effects models, the debiased inference for pooled regression, and consider an application to the Granger causality testing. In this paper, we explore how to use their framework for nowcasting large panels of low-frequency time series.

\medskip
	
We focus on nowcasting current quarter firm-specific price-earnings ratios (henceforth P/E ratios). This means we focus on evaluating model-based within-quarter predictions for very short horizons. It is widely acknowledged that P/E ratios are a good indicator of the future performance of a company and, therefore, are used by analysts and investment professionals to base their decisions on which stocks to pick for their investment portfolios. Typically investors rely on consensus forecasts of earnings made by a pool of analysts. We, therefore, choose such consensus forecasts as the benchmark for our proposed machine learning methods. \cite{ball2018automated} and \cite{carabias2018real} documented that analysts tend to focus on their firm/industry when making earnings predictions while not fully taking into account the impact of macroeconomic events.  \cite{babii2022machine} tested formally in a high-dimensional data setting the hypothesis that systematic and predictable errors occur in analyst forecasts and confirmed empirically that they {\it leave money on the table}. The analysis in the current paper is therefore an logical extension of this prior work.
In addition, we also compare our proposed new methods with the MIDAS regression forecast combination approach used by \cite{ball2018automated} as well as a simple random walk model.

\medskip

Our high-frequency regressors include traditional macro and financial series as well as non-standard series generated by textual analysis of financial news. We consider structured pooled and fixed effects sg-LASSO panel data regressions with mixed frequency data (sg-LASSO MIDAS). By ``structured'' we mean that the ML procedure is set up such that it recognizes the time series and panel structure of the data. This is a departure from standard ML which is rooted in a tradition of i.i.d.\ covariates and therefore time series and panel data structures are not recognized. For the purpose of comparison, we include elastic net estimators in our analysis, as a representative example of standard ML.

\medskip

In our empirical analysis we study nowcasting the firm-level P/E ratio for a large set of firms. Moreover, we decompose the (log of) the P/E ratio into the return for firm $i$ and analyst prediction errors. Therefore, nowcasting the log P/E ratio could also be achieved via nowcasting its two components. The decomposition corresponds to the distinction between analyst assessments of firm $i$'s earnings and market/investor assessments of the firm.

\medskip

Our empirical results can be summarized as follows.  Predictions based on analyst consensus exhibit significantly higher mean squared forecast errors (MSEs) compared to model-based predictions. These model-based predictions involve either direct log P/E ratio nowcasts or their individual components. The MSE for the random walk model and analysts' concensus are quite similar, and therefore random walk predictions are outperformed by the model-based ones as well. A substantial proportion of firms (approximately 60\%) exhibit low MSE values, indicating a high level of prediction accuracy. However, there are a few firms for which the MSEs are relatively larger, suggesting lower prediction performance for these specific cases. Comparing  direct log P/E ratio nowcasts versus those based on its components, we observe a substantial improvement in prediction accuracy when using the individual components. This improvement is consistently evident across individual, pooled, and fixed effects regression models. Moreover, the sparsity patterns differ significantly across the direct versus component prediction models. 

\medskip

Our framework allows us to go beyond providing quarterly nowcasts and generate daily updates of earnings series. Leveraging the daily influx of information throughout the quarter, we continuously re-estimate our models and produce nowcast updates as soon as new data becomes available. We report the distribution of Mean Squared Errors (MSEs) across firms for five distinct nowcast horizons: 20-day, 15-day, 10-day, and 5-day ahead, as well as the end of the quarter and show that as the horizons become shorter, both the median and upper quartile of MSEs decrease. 
The sg-LASSO estimator we employ in our study is well-suited for incorporating grouped fixed effects. This approach involves grouping firm-specific intercepts based on either statistical procedures or economic reasoning, as outlined in \cite{bonhomme2015grouped}. In our analysis, we utilize the Fama French industry classification to form 10 distinct groups for grouping fixed effects. Our findings suggest that grouped fixed effects strike a better balance between capturing heterogeneity and pooled parameters, resulting in more accurate nowcast predictions. These results support the notion that incorporating group fixed effects enhances the overall performance of our forecasting model.

\medskip

Next we address the challenge of missing earnings data, which can complicate the analysis. We examine the performance of parameter imputation methods in computing nowcasts, see, e.g, \cite{brown2023nowcasting}, even when earnings and/or earnings forecasts are missing for certain observations in the sample. 
The results obtained through parameter imputation outperform the analyst consensus nowcasts in terms of prediction accuracy. 

\medskip

The paper is organized as follows. Section \ref{sec:method} introduces the models and estimators. A simulation study reporting the finite sample nowcasting performance of our proposed methods appears in Section \ref{sec:mc}. The results of our empirical application analyzing price-earnings ratios for a panel of individual firms are reported in Section \ref{sec:emp}. Section \ref{sec:conclusion} concludes. All technical details and detailed data descriptions appear in the Appendix and the Online Appendix.

\section{High-dimensional mixed frequency panel data}\label{sec:method}
In this section, we describe the methodological approach of the paper. Motivated by our application, we will refer to the cross-sectional observations as firms, the low-frequency observations as quarterly while the high-frequency observations are daily or monthly. However, the notation presented in this section is generic and can correspond to other entities and frequencies. The objective is to nowcast  $\{y_{i,t}:i\in[N],t\in[T]\}$ (where for a positive integer $p$, we put $[p]=\{1,2,\dots,p\}$), in our case a panel of P/E ratios (or its decomposition into returns and analyst forecast errors) for $N$ firms observed at $T$ time periods. The covariates consist of $K$ time-varying predictors measured potentially at higher frequencies
\begin{equation*}
	\left\{x_{i,t-j/n_k^H,k}:\;i\in[N],t\in[T],j=0,\dots,n^L_kn_k^H-1,k\in[K]\right\},
\end{equation*}
where $n_k^H$ is the number of high-frequency observations for the $k^{\rm th}$ covariate in a low-frequency time period $t$, and $n^L_k$ is the number of low-frequency time periods used as lags. For instance, $n^L_k=1$ corresponds in our application to a quarter of high-frequency lags used as covariates and $n_k^H=3$ corresponds to monthly data with 3 month of data available per quarter. Note that we can think of mixtures of say annual, quarterly, monthly and weekly data, and therefore $n_k^H$ represents different high frequency sampling frequencies and associated lags $n_k^L n_k^H.$

\medskip

In our empirical analysis we examine three types of regression model specifications: (a) regularized single equation regressions for each individual firm, (b) regularized panel regressions with pooling, and (c) regularized panel regressions with fixed effects. Hence, in (a) we do not explore the panel structure of the data, whereas in (b) and (c) we do. To discuss the model specifications, we focus here on (b) and (c), keeping in mind that the single regression case is a straightforward simplification of the panel regression models.

\medskip

Consider the mixed frequency panel data regression for  $y_{i,t|\tau},$ that is observation $i$ for low-frequency nowcasting $y$ at time $t$ using information up to $\tau:$  
\begin{equation*}
	y_{i,t|\tau} = \alpha_i + \sum_{k=1}^{K}\psi(L^{1/n_k^H};\beta_k)x_{i,\tau,k} + u_{i,t|\tau},
\end{equation*}
where $\alpha_i$ is the entity-specific intercept (depending on $\tau$ but we suppress this detail to simplify notation), and
\begin{equation}\label{eq:hf_lag}
	\psi(L^{1/n_k^H};\beta_k)x_{i,\tau,k} = \frac{1}{k_{max}}\sum_{j=0}^{k_{max}-1}\beta_{j,k}L^{j/n_k^H}x_{i,\tau,k}
\end{equation}
where $k_{max}$ is the maximum lag length which may depend on the covariate $k,$ and for each high frequency covariate $x_{i,\tau,k}$ we have the most up to date information available at time $\tau.$ This may imply that for some high frequency regressors this is stale information as they have not been updated yet, but presumably at least some of the high frequency data are fresh real-time information at the time $\tau$ the nowcast is being made. 
For instance, in our quarterly/monthly application we can have $\tau$ = $(t - 1) + 1/3$ in which case we nowcast quarter $t$ with information available at the end of the first month of that quarter. In this example, some high frequency series for the first month may be available while some may not due to say publication lags. Likewise, with $\tau$ = $(t - 1) + 2/3$ we can revise the previous nowcast with one extra month of information, which taking into account publication lags may include observations from the first month as the most recent releases. It should parenthetically be noted that for $\tau$ $\leq$ $t - 1,$ we are dealing with a forecasting situation and therefore our analysis applies to both nowcasting and - ceteris paribus - forecasting.

\medskip

To reduce the dimensionality of the high-frequency lag polynomial, we follow the MIDAS ML literature, see \cite{babii2020inference,babii2020machine}, and estimate a weight function $\omega$ parameterized by a relatively small number of coefficients $L$
\begin{equation*}
		\psi(L^{1/n_k^H};\beta_k)x_{i,\tau,k} = 
	\frac{1}{k_{max}}\sum_{j=0}^{k_{max}-1}\omega\left(\frac{j}{n_k^H};\beta_k\right)x_{i,\tau,k},
\end{equation*}
where the MIDAS weight function is
$\omega(s;\beta_k)$ = $\sum_{l=0}^{L-1}\beta_{l,k}w_l(s),$ 
$(w_l)_{l\geq 0}$ is a collection of $L$ approximating functions, called the \textit{dictionary}, and $\beta_k\in\R^L$ is the unknown parameter. An example of a dictionary used in the MIDAS ML literature is the set of orthogonal Legendre polynomials. To streamline notation it will be convenient to assume, without loss of generality, a common lag length, i.e.\ $\bar{k}_{max}$ = $k_{max}$ $\forall$ $k$ $\in$ $[K].$
The linear in parameters dictionaries map the MIDAS regression to a standard linear regression framework. In particular,  define $\mathbf{x}_i = (X_{i,1}W,\dots,X_{i,K}W)$, where for each $k\in[K]$, $X_{i,k} = (x_{i,\tau-j/n_k^H,k},j = 0, \ldots, \bar{k}_{max} - 1)_{\tau\in[T]}$ is a $T\times \bar{k}_{max}$ matrix of covariates and $\bar{k}_{max} W$ = $(w_l(j/n_k^H; \beta_k)_{0\leq l\leq L-1, 0 \leq j\leq \bar{k}_{max}}$ is a $\bar{k}_{max} \times L$ matrix corresponding to the dictionary. In addition, let 
$\mathbf{y}_i$ = $(y_{i,t|\tau}, t, \tau \in [T])^\top$ and $\mathbf{u}_i$ = $(u_{i,t|\tau},t, \tau \in [T])^\top.$
The regression equation after stacking time series observations for each firm $i\in[N]$ is as follows
\begin{equation*}
	\mathbf{y}_i = \iota\alpha_i + \mathbf{x}_i\beta + \mathbf{u}_i,
\end{equation*}
where $\iota\in\R^T$ is the all-ones vector and $\beta\in\R^{LK}$ is a vector of slope coefficients. Lastly, put $\mathbf{y} = (\mathbf{y}_1^\top,\dots, \mathbf{y}_N^\top)^\top$, $\mathbf{X}=(\mathbf{x}_1^\top, \dots, \mathbf{x}_N^\top)^\top$, and $\mathbf{u} = (\mathbf{u}_1^\top,\dots,\mathbf{u}_N^\top)^\top$. Then the regression equation after stacking all cross-sectional observations is
\begin{equation*}
\mathbf{y} = B\alpha + \mathbf{X}\beta + \mathbf{u},
\end{equation*}
where  $B=I_N\otimes\iota$, $I_N$ is $N\times N$ identity matrix, and $\otimes$ is the Kronecker product.
Given that the number of potential predictors $K$ can be large, additional regularization can improve the predictive performance in small samples. To that end, we take advantage of the sg-LASSO regularization, suggested by \cite{babii2020machine}. 

\medskip

The fixed effects sg-LASSO estimator $\hat\rho=(\hat\alpha^\top,\hat\beta^\top)^\top$ solves
\begin{equation}\label{eq:sgl}
	\min_{(a,b)\in\R^{N+p}}\|\mathbf{y} - Ba - \mathbf{X}b \|_{NT}^2 + 2 \lambda\Omega(b),
\end{equation}
where $\Omega$ is the sg-LASSO regularizing functional. It is worth stressing that the design matrix $\mathbf{X}$ does not include the intercept and that we do not penalize the fixed effects which are typically not sparse. In addition,  $\|.\|_{NT}^2 = |.|^2/(NT)$ is the empirical norm and
\begin{equation*}
	\Omega(b) = \gamma|b|_1 + (1-\gamma)\|b\|_{2,1},
\end{equation*}
is a regularizing functional. It is a linear combination of the $\ell_1$ LASSO and $\ell_{2,1}$ group LASSO norms. Note that for a group structure $\mathcal{G}$ described as a partition of $[p]=\{1,2,\dots,p\}$, the group LASSO norm is computed as $\|b\|_{2,1}=\sum_{G\in\mathcal{G}}|b_G|_2$, while $|.|_q$ denotes the usual $\ell_q$ norm. The group LASSO penalty encourages sparsity between groups whereas the $\ell_1$ LASSO norm  promotes sparsity within groups and allows us to learn the shape of the MIDAS weights from the data.   The parameter $\gamma\in[0,1]$ determines the relative weights of the $\ell_1$ (sparsity) and the $\ell_{2,1}$ (group sparsity) norms, while the amount of regularization is controlled by the regularization parameter $\lambda\geq 0$. 

\medskip

In Section~\ref{sec:intro}, we called our approach structured ML because the group structure allows us to embed the time series structure of the data. More specifically, these structures are represented by groups covering lagged dependent variables and groups of lags for a single (high-frequency) covariate. 
Throughout the paper, we assume that groups have fixed size,  and the group structure is known by the econometrician. Both are reasonable assumptions to make in the context of our empirical application.



\medskip

For pooled regressions, we assume that all entities share the same intercept parameter $\alpha_1=\dots=\alpha_N=\alpha$. The pooled sg-LASSO estimator $\hat\rho=(\hat\alpha,\hat\beta^\top)^\top$ solves
\begin{equation}\label{eq:pooled_panel}
	\min_{r=(a,b)\in\R^{1+p}}\|\mathbf{y} - a\iota - \mathbf{X}b\|_{NT}^2 + 2 \lambda\Omega(r).
\end{equation}
Pooled regressions are attractive since the effective sample size $NT$ can be huge, yet the heterogeneity of individual time series may be lost. If the underlying series have a substantial heterogeneity over $i\in[N]$, then taking this into account might reduce the projection error and improve the predictive accuracy.
 
\medskip

\cite{babii2022machine} provide the theoretical analysis of predictive performance of regularized panel data regressions with the sg-LASSO regularization, including as special cases (a) standard LASSO, (b) group LASSO regularizations as well as (c) generic high-dimensional panels not involving mixed frequency data. Finally,
\cite{babii2022machine} also develop the debiased inferential methods and Granger causality tests for pooled panel data regressions.

\section{Monte Carlo experiments \label{sec:mc}}

It is not clear that the aforementioned theory is of practical use in the context of nowcasting using modestly sized samples of data. For this reason,  we investigate in this section the finite sample nowcasting performance of the machine learning methods covered so far. We consider the standard (unstructured) elastic net with UMIDAS (called Elnet-U), where UMIDAS refers to unconstrained MIDAS proposed by \cite{FMS15} in a classic non-ML context, and sg-LASSO with MIDAS. Both methods require selecting two tuning parameters $\lambda$ and $\gamma$. In the case of sg-LASSO, $\gamma$ is the relative weight of LASSO and group LASSO penalties while in the case of the elastic net $\gamma$ interpolates between LASSO and ridge. In both cases we report results on a grid $\gamma \in \{0, 0.2, \dots, 1\}$.

\medskip

In addition to evaluating the performance over the grid of $\gamma$ tuning parameter values, we need to select the $\lambda$  tuning parameter. To do so, we consider several approaches. First, we adapt the $K$-fold cross-validation to the panel data setting. To that end, we resample the data by blocks respecting the time-series dimension and creating folds based on cross-sectional units instead of the pooled sample. We use the 5-fold cross-validation both in the simulation experiments and the empirical application. We also consider the following three information criteria: BIC, AIC, and corrected AIC (AICc) of \cite{hurvich1989regression}. Assuming that $y_{i,t}|x_{i,t}$ are i.i.d.\ draws from $N(\alpha_i + {x}_{i,t}^\top\beta, \sigma^2)$, the log-likelihood of the sample is
\begin{equation*}
	\mathcal{L}(\alpha,\beta,\sigma^2) \propto - \frac{1}{2\sigma^2}\sum_{i=1}^N\sum_{t=1}^T(y_{i,t} - \alpha_i - x_{i,t}^\top\beta)^2.
\end{equation*}
Then, the BIC criterion is
\begin{equation*}
	\mathrm{BIC} = \frac{\|\mathbf{y} - \hat\mu - \mathbf{X}\hat\beta \|_{NT}^2}{\hat\sigma^2} + \frac{\log(NT)}{NT}\times{df},
\end{equation*}
where $df$ denotes the degrees of freedom, $\hat\sigma^2$ is a consistent estimator of $\sigma^2$, $\hat\mu=\hat\alpha\iota$ for the pooled regression, and $\hat\mu=B\hat\alpha$ for fixed effects regression. The degrees of freedom are estimated as $\widehat{df} = |\hat\beta|_0+1$ for the pooled regression and $\widehat{df} = |\hat\beta|_0+N$ for the fixed effects regression, where $|.|_0$ is the $\ell_0$-norm defined as a number of non-zero coefficients; see \cite{zou2007degrees} for more details. The AIC is computed as
\begin{equation*}
	\mathrm{AIC} = \frac{\|\mathbf{y} - \hat\mu - \mathbf{X}\hat\beta \|_{NT}^2}{\hat\sigma^2} + \frac{2}{NT}\times\widehat{df},
\end{equation*}
and the corrected Akaike information criteria is
\begin{equation*}
	\mathrm{AICc} = \frac{\|\mathbf{y} - \hat\mu - \mathbf{X}\hat\beta \|_{NT}^2}{\hat\sigma^2} + \frac{2\widehat{df}}{NT - \widehat{df} - 1}.
\end{equation*}
The AICc is typically a better choice when $p$ is large relative to the sample size. We report the results for each of the tuning parameter selection criteria for $\lambda,$ along the grid choice for $\gamma.$

\subsection{Simulation Design \label{appsec:mcdesign}}

To assess the predictive performance of pooled panel data models, we simulate the data from the following DGP with a quarterly/monthly frequency mix in mind and  $\bar{k}_{max}$ = $k_{max}$  with $n_k^H$ = $n^H$ $\forall$ $k:$
\begin{equation*}
	y_{i,t|\tau} = \alpha + \sum_{k=1}^K \bar{k}_{max}^{-1}\sum_{j=0}^{\bar{k}_{max}-1} \omega(j/n^H;\beta_k)x_{i,\tau-j/n^H,k} + u_{i,t|\tau},
\end{equation*} 
where $i\in[N]$, $t\in[T]$, $\alpha$ is the common intercept, $\bar{k}_{max}^{-1}\sum_{j=0}^{\bar{k}_{max}-1} \omega(j/n_k;\beta_k)$ the weight function for $k$-th high-frequency covariate and the error term is either  $u_{i,t|\tau} \sim_{i.i.d.}N(0,1)$ or $u_{i,t|\tau}  \sim_{i.i.d.}\text{student-}t(5)$. 

\medskip

We are interested in a quarterly/monthly data mix, and use four quarters of data for the high-frequency regressors which covers 12 high-frequency lags for each regressor. In terms of information sets we start with $\tau$ = $t - 1,$ which corresponds to a prediction setting and then have $\tau$ = $t - 1 + 1/3,$ i.e.\ nowcasting with one month's worth of information. 
We set the number of relevant high-frequency regressors $K$ = 6. The high-frequency regressors are generated as  $K$ i.i.d.\ realizations of the univariate autoregressive (AR) process $x_h = \rho x_{h-1}+ \varepsilon_h,$ where $\rho=0.6$ and either $\varepsilon_h\sim_{i.i.d.}N(0,1)$ or $\varepsilon_h\sim_{i.i.d.}\text{student-}t(5)$, where $h$ denotes the high-frequency sampling. We rely on a commonly used weighting scheme in the MIDAS literature, namely $\omega(s;\beta_k)$ for $k=1,2,\dots,6$ are determined by beta densities respectively equal to $\mathrm{Beta}(1,3)$ for $k=1,4$, $\mathrm{Beta}(2,3)$ for $k=2,5$, and $\mathrm{Beta}(2,2)$ for $k=3,6$; see \cite{ghysels2007midas} or \cite{ghysels2019estimating}, for further details. The MIDAS regressions are estimated using Legendre polynomials of degree $L=3$. 

\medskip

We consider DGPs featuring pooled panels and fixed effects. For the pooled panel regression DGPs we simulate the intercepts as $\alpha \sim \text{Uniform}(-4,4).$ For the fixed effects models 
the individual fixed effects are simulated as $\alpha_i \sim_{\text{i.i.d}} \text{Uniform}(-4,4)$ and are kept fixed throughout the experiment. 

\medskip

For $\tau$ = $t - 1,$ the {\it Baseline scenario}, in the estimation procedure we add 24 noisy covariates which are generated in the same way as the relevant covariates, use 4 low-frequency lags and the error terms $u_{i,t|\tau}$ and $\varepsilon_h$ are Gaussian. In the student-$t(5)$ scenario we replace the Gaussian error terms with a student-$t(5)$ distribution while in the {\it large dimensional} scenario we add 94 noisy covariates. For each scenario, we simulate $N=25$ i.i.d.\ time series of length $T=50$; next we increase the cross-sectional dimension to $N=75$ and time series to $T=100$. 

\medskip

Finally, for $\tau$ = $t - 1 + 1/3$ the thought experiment in the simulation design is one where the first high-frequency observations during low frequency $t$ are available. The nowcaster of course does not know which of the covariates are relevant nor does she know the parameters of the prediction rule. We will call this scheme ``one-step ahead'' nowcasts. 

\subsection{Simulation results}

Tables \ref{main_mcres} and \ref{main_mcres2}  cover the average mean squared forecast errors (MSFE) for one-step ahead nowcasts for the three simulation scenarios. We report results for sg-LASSO with MIDAS weights (left block) and elastic net with UMIDAS (right block) using both pooled panel models (Table \ref{main_mcres}) and fixed effects ones (Table \ref{main_mcres2}). We report results for the best choice of the $\gamma$ tuning parameter.\footnote{Results for the grid of $\gamma\in\{0.0, 0.2, \dots, 1.0\}$ are reported in the Online Appendix Tables \ref{app:main_mcres1}-\ref{app:main_mcres3}.}

\begin{table}[!htbp]
	\centering
	\small{
		\setlength{\tabcolsep}{4.5pt}
		\begin{tabular}{c rrr  r rrr}
			\rule{0pt}{4ex} 
			&\multicolumn{3}{c}{\underline{sg-LASSO}} &&\multicolumn{3}{c}{\underline{Elnet-U}}\\
			\rule{0pt}{4ex} 
			N/T = & 25/50  & 75/50 & 25/100   & & 25/50  & 75/50 & 25/100  \\
			\hline
			\rule{0pt}{3ex} 
			&\multicolumn{7}{c}{\underline{Panel A. Baseline}} \\
			CV &1.191 &1.157&1.168&&1.213 &1.158 & 1.172 \\
			BIC &1.270 & 1.175 &1.202 &&1.384 & 1.211 & 1.247\\
			AIC &1.234 & 1.160 & 1.187 &&1.273&1.172 &  1.213\\
			AICc &1.237 & 1.161 & 1.188 &&1.279&1.172&1.217\\
			&\multicolumn{7}{c}{\underline{Panel B. Student-$t(5)$}} \\
			CV & 1.280 & 1.245 & 1.248 &&1.299&1.243&1.256\\
			BIC &1.389 & 1.274 & 1.293 &&1.570&1.317&1.367\\
			AIC & 1.345 &1.259 & 1.272 &&1.411&1.283&1.298\\
			AICc & 1.344&1.259&1.273&&1.412&1.283&1.300\\
			&\multicolumn{7}{c}{\underline{Panel C. Large-dimensional}} \\
			CV & 1.204 &1.160&1.185&&1.255&1.165&1.188\\
			BIC &1.273 &1.175&1.214&&1.409&1.208&1.289\\
			AIC & 1.259&1.166&1.191&&1.350&1.198&1.232\\
			AICc & 1.260&1.167&1.192&&1.353&1.200&1.232\\
			\hline\hline
		\end{tabular}
	}
	\caption{\small The table reports the MSFE for nowcasting accuracy for the pooled estimator for the Baseline (Panel A), student-$t(5)$ (Panel B), and  large-dimensional (Panel C) DGPs for the sg-LASSO-MIDAS (rows sg-LASSO) and elastic net UMIDAS (rows Elnet-U). We vary the cross-sectional dimension $N\in\{25, 75\}$ and time series dimension $T\in\{50, 100\}$.  We report results for 5-fold cross-validation, BIC, AIC, AICc information criteria $\lambda$ tuning parameter calculation methods and for the best choice of $\gamma$ tuning parameter. \label{main_mcres}} 
\end{table}

\begin{table}[htbp]
	\centering
	\small{
		\setlength{\tabcolsep}{4.5pt}
		\begin{tabular}{c rrr  r rrr}
			\rule{0pt}{4ex} 
			&\multicolumn{3}{c}{\underline{sg-LASSO}} &&\multicolumn{3}{c}{\underline{Elnet-U}}\\
			\rule{0pt}{4ex} 
			N/T = & 25/50  & 75/50 & 25/100   & & 25/50  & 75/50 & 25/100  \\
			\hline
			\rule{0pt}{3ex} 
			&\multicolumn{7}{c}{\underline{Panel A. Baseline}} \\
			CV & 1.198 &1.170&1.164 & &1.245&1.183&1.184\\
			BIC & 1.304 & 1.202 &1.213 &&1.537&1.259&1.313\\
			AIC & 1.282&1.192&1.196&&1.380&1.222&1.237\\
			AICc &1.284 &1.193&1.196&&1.284&1.193&1.196\\
			&\multicolumn{7}{c}{\underline{Panel B. Student-$t(5)$}} \\
			CV & 1.278 &1.256 &1.248 &&1.329 &1.270&1.271\\
			BIC & 1.437 &1.306 &1.310&&1.694&1.367&1.404\\
			AIC & 1.389&1.292&1.294&&1.478&1.316&1.342\\
			AICc & 1.393&1.293&1.295&&1.495&1.316&1.348\\
			&\multicolumn{7}{c}{\underline{Panel C. Large-dimensional}} \\
			CV & 1.214 &1.170&1.172&&1.282&1.197&1.193\\
			BIC & 1.344&1.213&1.229&&1.662&1.298&1.342\\
			AIC &1.300 &1.243&1.202&&1.404&1.384&1.235\\
			AICc & 1.301&1.205&1.204&&1.405&1.247&1.238\\
			\hline\hline
		\end{tabular}
	}
	\caption{\small The table reports the MSFE for nowcasting accuracy for the fixed effects estimator for the Baseline (Panel A), student-$t(5)$ (Panel B), and  large-dimensional (Panel C) DGPs for the sg-LASSO-MIDAS (rows sg-LASSO) and elastic net UMIDAS (rows Elnet-U). We vary the cross-sectional dimension $N\in\{25, 75\}$ and time series dimension $T\in\{50, 100\}$.  We report results for 5-fold cross-validation, BIC, AIC, AICc information criteria $\lambda$ tuning parameter calculation methods and for the best choice of $\gamma$ tuning parameter. \label{main_mcres2}} 
\end{table}

Firstly, structured sg-LASSO-MIDAS consistently outperforms unstructured Elnet-U for all DGPs and in both pooled and fixed effects cases. The most significant discrepancy between the two methods is observed in situations with small N and small T, specifically when N = 25 and T = 50. As either N or T increases, this gap gradually diminishes. When comparing the results of pooled and fixed effects, it becomes evident that the difference between the two approaches — structured sg-LASSO-MIDAS versus Elnet UMIDAS — widens further in the case of fixed effects with student-t(5) data. This indicates that our structured approach yields higher quality estimates for the fixed effects and thus more accurate nowcasts.

\medskip

In the case of sg-LASSO-MIDAS, the best performance is achieved for $\gamma\notin\{0,1\}$ for both pooled panel data and fixed effects cases, while $\gamma=0$, i.e.\ ridge regression, seems to be dominated by estimators that $\gamma\notin\{0,1\}$  in both pooled and fixed effects cases. For the student-$t(5)$ and large dimensional DGP, we observe a decrease in the performance for all methods. However, the decrease in the performance is larger for the student-$t(5)$ DGP, revealing that heavy-tailed data have  —  as expected  —  a stronger impact on the performance of the estimators. 

\medskip

For the pooled panel data case, increasing $N$ from $25$ to $75$ seems to have a larger positive impact on the performance than an increase in the time-series dimension from $T=50$ to $T=100$. The difference appears to be larger for student-$t(5)$ and large dimensional DGPs and/or for the elastic net case. Turning to the fixed effects results, the differences seem to be even sharper, in particular for student-$t(5)$ and large dimensional DGPs.

\medskip

When comparing the results across the different model selection methods, i.e., cross-validation and the three information criteria, we find that almost always cross-validation leads to smaller prediction errors in both pooled and fixed effects panel data cases. Notably, the gains appear to be larger for the large $N$ and $T$ values. Comparing BIC, AIC, and AICc information criteria, the results appear to be similar for AIC and AICc across DGPs and different sample sizes, while the BIC performance is slightly worse than AIC and AICc.

\section{Nowcasting price-earnings ratios \label{sec:emp}}

\cite{ball2018automated}, \cite{carabias2018real} and \cite{babii2022machine}  documented that analysts make systematic and predictable errors in their P/E forecasts. 
We therefore consider nowcasting the P/E ratios using a set of predictors that are sampled at mixed frequencies for a large cross-section of firms.

\medskip

A natural question one may ask: should we nowcast P/E ratio directly or it's components. We, therefore, decompose the (log of) the P/E ratio for firm $i$ as follows:
\begin{eqnarray}
	\label{eq:decomp1}
	pe_{i,t+1} 	\equiv \log (P_{i,t+1}/E_{i,t+1}) & = &  	\log ((P_{i,t+1}/P_{i,t})/(E_{i,t+1}/P_{i,t})) \notag \\
	& = & r_{i,t+1} - 	\log ((E_{i,t+1}/E^a_{i,t+1|t})/(P_{i,t}/E^a_{i,t+1|t})) \notag \\
	& = & r_{i,t+1} - e^a_{i,t+1|t} +	\log (P_{i,t}/E^a_{i,t+1|t}) 
\end{eqnarray}
where $r_{i,t+1}$ is the log return from $t + 1$ to $t$ for firm $i,$ $E^a_{i,t+1|t}$ the analyst's prediction at time $t$ pertaining to $t + 1$ earnings, and $e^a_{i,t+1|t}$ $\equiv$ $\log (E_{i,t+1}) - \log (E^a_{i,t+1|t})$ is the log earnings forecast error of analysts pertaining to their end of period $t$ prediction for $t + 1.$ Finally, $\log (P_{i,t}/E^a_{i,t+1|t})$ is perfectly known at time $t.$ The above defines an additive decomposition of the log P/E ratio into the return for firm $i$ and the analyst prediction error. Therefore, nowcasting the log P/E ratio could also be achieved via nowcasting its two components. The decomposition corresponds to the distinction between analyst assessments of firm $i$'s earnings and market/investor assessments of the firm.

\medskip

There is a considerable literature on using machine learning to predict returns, see e.g.\  \cite{rapach2010out}, \cite{kim2014forecasting}, \cite{gu2020empirical}, \cite{d2020artificial}, among others. Here we are dealing with a slightly modified setting where we are nowcasting quarterly returns with information during quarter $t + 1.$ Nevertheless, prediction and nowcasting are closely related. The second component, $e^a_{i,t+1|t}$ has been explored by \cite{babii2022machine}, who revisit a topic raised by \cite{ball2018automated} and \cite{carabias2018real}, and confirmed in a rich data setting that analysts tend to focus on their firm/industry when making earnings predictions while not fully taking into account the impact of macroeconomic events. Put differently, one can forecast and nowcast analyst prediction errors. 

\medskip

It should also parenthetically be noted that equation (\ref{eq:decomp1}) can be rewritten as a decomposition of returns, namely:
\begin{equation}
	\label{eq:decomp2}
	r_{i,t+1}  = 	pe_{i,t+1} +  e^a_{i,t+1|t} + \log (P_{i,t}/E^a_{i,t+1|t}) 
\end{equation}
which can be viewed as an alternative decomposition of returns compared to \cite{ferreira2011forecasting}. They propose forecasting separately the three components of stock market returns: (a) the dividend price ratio, (b) earnings growth, and (c) price-to-earnings ratio growth. \cite{ferreira2011forecasting} argue that predicting the separate components yields better return predictions compared to the usual models producing direct forecasts of the latter. They  estimate the expected earnings growth using a 20-year moving average of the growth in earnings per share. The expected dividend price ratio is estimated by the current dividend price ratio. This implicitly assumes that the dividend price ratio follows a random walk. While our application is different in many regards, the arguments being considered are similar. It is worth reminding ourselves that if the nowcast $\widehat{pe}_{i,t+1}$ is constructed from individual component nowcasts, then
\begin{equation}
	\label{eq:MSEcomponents}
	\text{MSE}(\widehat{pe}_{i,t+1}) = \text{MSE}(\widehat r_{i,t+1}) + \text{MSE}(\hat e^a_{i,t+1|t}) - 2\E\left[(r_{i,t+1} - \hat r_{i,t+1})(e^a_{i,t+1|t} - \hat e^a_{i,t+1|t})\right]
\end{equation}
Hence, depending on the co-movements between returns for firm $i,$ $r_{i,t+1}$ and analyst earning prediction errors $e^a_{i,t+1|t},$ we are better off to directly predict $pe_{i,t+1}$ or its components. If the latter are positively correlated, then we are better off direct forecasting is preferred.

\medskip

Given the aforementioned decomposition, we are interested in the following LHS variables: $pe_{i,t+1},$ $r_{i,t+1}$ and  $e^a_{i,t+1|t}.$ 
First, we estimate the individual sg-LASSO MIDAS regressions for each firm $i=1,\dots,N$, namely:
\begin{equation*}
	\mathbf{y}_i = \iota\alpha_i + \mathbf{x}_i\beta_i + \mathbf{u}_i,
\end{equation*}
where the firm-specific predictions are computed as $\hat y_{i,t+1} = \hat\alpha_i + x_{i,t+1}^\top\hat\beta_i$. As noted in Section \ref{sec:method}, $ \mathbf{x}_i$ contains lags of the low-frequency target variable and high-frequency covariates to which we apply Legendre polynomials of degree $L=3$.

\medskip

Next, we estimate the following pooled and fixed effects sg-LASSO MIDAS panel data models
\begin{eqnarray*}
	\mathbf{y} & =  \alpha\iota + \mathbf{X}\beta + \mathbf{u} & \text{Pooled} \\
	\mathbf{y} & =  B\alpha + \mathbf{X}\beta + \mathbf{u} & \text{Fixed Effects} 
\end{eqnarray*}
and compute predictions as 
\begin{eqnarray*}
	\hat y_{i,t+1} & = \hat\alpha + x_{i,t+1}^\top\hat\beta & \text{Pooled} \\
	\hat y_{i,t+1} & = \hat\alpha_i + x_{i,t+1}^\top\hat\beta & \text{Fixed Effects}.
\end{eqnarray*}
Once we compute the forecast for the log of P/E ratio ($pe_{i,t+1}$), log returns ($r_{i,t+1}$) and log earnings forecast error ($e^a_{i,t+1|t}$), we compute the final prediction accuracy metrics by either taking directly log P/E nowcast or the sum of its components, i.e., $\hat{S} =  \hat r_{i,t+1} - \hat e^a_{i,t+1|t} +	\log (P_{i,t}/E^a_{i,t+1|t})$.

\smallskip 

We benchmark firm-specific and panel data regression-based nowcasts against two simple alternatives. First, we compute forecasts for the RW model as 
\begin{equation*}
	\hat y_{i,t+1|t} = y_{i,t}.
\end{equation*}
Second, we consider predictions of P/E implied by analysts' earnings nowcasts using the information up to time $t+1$, i.e.\ 
\begin{equation*}
	\hat y_{i,t+1|t} = \bar y_{i,t+1|t}^a,
\end{equation*}
where the predicted/nowcasted log of P/E ratio is based on consensus earnings forecasts pertaining to the end of the $t+1$ quarter using the stock price at the end of quarter $t.$ 
To measure the forecasting performance, we compute the mean squared forecast errors (MSE) for each method. Let $\mathbf{\bar y}_i = (y_{i,T_{is}+1},\dots, y_{i,T_{os}})^\top$ represent the out-of-sample realized P/E ratio values, where $T_{is}$ and $T_{os}$ denote the last in-sample observation for the first prediction and the last out-of-sample observation respectively, and let $\mathbf{\hat y}_i = (\hat y_{i,t_{is}+1},\dots, \hat y_{i,t_{os}})$ collect the out-of-sample forecasts. Then, the mean squared forecast errors are computed as
\begin{equation*}
	\mathrm{MSE} = \frac{1}{N} \sum_{i=1}^{N} \frac{1}{T-T_{is}+1} (\mathbf{\bar y}_i-\mathbf{\hat y}_i)^\top  (\mathbf{\bar y}_i-\mathbf{\hat y}_i).
\end{equation*}

\medskip

We look at 210 US firms and use 24 predictors, including traditional macro and financial series as well as non-traditional series from textual analysis of financial news. We apply (a) single regression individual firm high-dimensional regressions, (b) pooled and (c) individual fixed effects sg-LASSO MIDAS panel data models and report results for several choices of the tuning parameters. We compare these three type of models with several benchmarks, which include a random walk (RW) model and analysts' consensus forecasts. 
The remainder of the section is structured as follows. We start with a short review of the data followed by a summary of the empirical results.

\subsection{Data description}\label{sec:data}

The full sample consists of observations between the \(1^{st}\) of January, 2000 and the \(30^{th}\) of June, 2017. Due to the lagged dependent variables in the models, our effective sample starts in the third fiscal quarter of 2000. We use the first 25 observations for the initial sample, and use the remaining 42 observations for evaluating the out-of-sample forecasts, which we obtain by using an expanding window forecasting scheme. We collect data from CRSP and I/B/E/S  to compute the quarterly P/E ratios and firm-specific financial covariates; RavenPack is used to compute daily firm-level textual-analysis-based data; real-time monthly macroeconomic series are from the FRED-MD dataset, see \cite{mccracken2016fred} for more details; FRED is used to compute daily financial markets data and, lastly, monthly news attention series extracted from the {\it Wall Street Journal} articles are retrieved from \cite{bybee2019structure}.\footnote{The dataset is publicly available at \href{http://www.structureofnews.com/}{http://www.structureofnews.com/}.} Online Appendix Section \ref{appsec:data-description} provides a detailed description of the data sources.\footnote{In particular, firm-level variables, including P/E ratios, are described in Online Appendix Table \ref{tab:firm_data}, and the other predictor variables in Online Appendix Table \ref{tab:other_data}. The list of all firms we consider in our analysis appears in Online Appendix Table \ref{tab:list_firms}.}

\medskip

Our target variable is the P/E ratio for each firm. To compute it, we use CRSP stock price data and I/B/E/S earnings data. Earnings data are subject to release delays of 1 to 2 months depending on the firm and quarter. Therefore, to reflect the real-time information flow, we compute the target variable using stock prices that are available in real-time. We also take into account that different firms have different fiscal quarters, which also affects the real-time information flow. 

\medskip

For example, suppose for a particular firm the fiscal quarters are at the end of the third month in a quarter, i.e.\ end of March, June, September, and December. The consensus forecast of the P/E ratio is computed using the same end-of-quarter price data which is divided by the earnings consensus forecast value. The consensus is computed by taking all individual prediction values up to the end of the quarter and aggregating those values by taking either the mean or the median. To compute the target variable, we adjust for publication lags and use prices of the publication date instead of the end of fiscal quarter prices. More precisely, suppose we predict the P/E ratio for the first quarter. As noted earlier, earnings are typically published with 1 to 2 months delay; say for a particular firm the data is published on the 25$th$ of April. In this case, we record the stock price for the firm on 25$th$ of April, and divide it by the earnings announced on that date. 

\subsection{Models and main results}

To simplify the exposition, we denote $y$ as one of the three target variables we consider. 
The main findings from our analysis are presented in Table \ref{tab:nowcasts}. 
Column $\hat pe_{i,t+1}$ reports results for directly nowcasting the log P/E ratio, column $\hat S$ reports the results of nowcasting and summing up the components, column $r_{i,t+1}$ reports results for the log return component and column $\hat e^a_{i,t+1|t}$ reports results for the log earnings forecast error of analysts component. Row {\it RW} reports results for the random walk, while row {\it Consensus} for the median consensus nowcast. Panels {\it Individual},  {\it Pooled} and {\it Fixed effects} report results for different panel data models relative to the consensus MSE (columns $\hat pe_{i,t+1}$ and $\hat S$) and for the components (columns $r_{i,t+1}$ and $\hat e^a_{i,t+1|t}$) we report ratios
relative to the RW MSE since there are obviously no concensus series notably for the analyst forecast errors. 

\begin{center}
	{\it Nowcasting Performance}
\end{center}

In light of the simulation evidence, we report the empirical results using cross-validation in Table \ref{tab:nowcasts} and provide the full set of results in Online Appendix Table \ref{apptab:nowcasts}.
The entries in the top panel of Table \ref{tab:nowcasts} reveal that predictions based on analyst consensus exhibit significantly higher mean squared forecast errors (MSEs) compared to model-based predictions since all the ratios with respect to the concensus are less than one (see first two columns). These model-based predictions involve either direct log P/E ratio nowcasts (first column) or their individual components (second column). Since the MSE for RW and concensus are quite similar, this also implies that RW predictions are outperformed by the model-based ones. The substantial improvement in the accuracy of model-based predictions compared to analyst-based predictions underscores the value of employing machine learning techniques for nowcasting log P/E ratios. Across various machine learning methods, including single-firm and panel data regressions, we consistently observe enhanced performance.
\begin{table}
	\centering
	{\small 
	\begin{tabular}{lccccc}
		& & & & & \\
		& $\hat pe_{i,t+1}$ & $\hat S$ &	& $\hat r_{i,t+1}$ & $ \hat e^a_{i,t+1|t}$\\
		& & & & & \\ 
		& \multicolumn{5}{c}{All firms} \\
				& & & & & \\ 
		RW & 1.355  & & & 0.054 & 0.194 \\ 
		Consensus& 1.305  & & & & \\ 
		&\multicolumn{5}{c}{\textit{Individual}} \\ 
		 & 0.905 & 0.890 & & 1.088 & 0.848\\ 
		DM p-val RW & 0.117 & 0.115 & & 0.181 & 0.090\\ 
		DM p-val Cons. & 0.156 & 0.131 & &  &\\ 
		&\multicolumn{5}{c}{\textit{Pooled}} \\ 
		 & 0.894 & 0.790 & & 0.964 & 0.799  \\ 
		  DM p-val RW & 0.060 & 0.023 & & 0.128 & 0.021 \\ 
		  DM p-val Cons. & 0.075 & 0.053 & &  &\\ 
		&\multicolumn{5}{c}{\textit{Fixed effects}} \\
		 & 0.814 & 0.793 & & 0.971 & 0.803 \\ 
		  DM p-val RW & 0.051 & 0.033 & & 0.164  & 0.032 \\ 
		  DM p-val Cons. & 0.078 & 0.063 & &  &\\ 
		& & & & & \\ 
			& \multicolumn{5}{c}{	With single CCI outlier removed (see Figure \ref{fig:hist})}\\
					& & & & & \\ 
RW & 1.333  & & &  0.053 &0.173  \\ 
Consensus& 1.275  & & & & \\ 
&\multicolumn{5}{c}{\textit{Individual}} \\ 
& 0.978 & 0.790 & & 1.001 & 0.812\\ 
DM p-val RW &  0.585  & 0.027& &  0.912 & 0.081 \\ 
DM p-val Cons. &0.606 &0.034 & &  &\\ 
&\multicolumn{5}{c}{\textit{Pooled}} \\ 
& 0.777 & 0.768& &  0.943 & 0.788  \\ 
DM p-val RW & 0.025 & 0.004 & & 0.103 & 0.018 \\ 
DM p-val Cons. & 0.029 & 0.006 & &  &\\ 
&\multicolumn{5}{c}{\textit{Fixed effects}} \\
& 0.782 &  0.767 & & 0.954 & 0.783 \\ 
DM p-val RW & 0.028& 0.004 & & 0.119  & 0.021  \\ 
DM p-val Cons. & 0.030  & 0.006 & &  &\\ 
\hline
	\end{tabular}}
	\caption{\footnotesize Column $\hat pe_{i,t+1}$ reports results for directly nowcasting the log P/E ratio, $\hat S$ for nowcasting and summing up the components,  $r_{i,t+1}$ for the log return and $\hat e^a_{i,t+1|t}$ for the log earnings forecast error of analysts. {\it RW} is for the random walk, while {\it Consensus} is the median consensus nowcast. Panels {\it Individual},  {\it Pooled} and {\it Fixed effects} report results for models relative to the consensus MSE ($\hat pe_{i,t+1}$ and $\hat S$) and for the components ($r_{i,t+1}$ and $\hat e^a_{i,t+1|t}$) relative to the RW MSE. DM is the \cite{diebold1995comparing} test statistic p-values using one-sided critical values. \label{tab:nowcasts}
}
\end{table}
When comparing the first and second columns, which correspond to direct log P/E ratio nowcasts versus those based on its components, we observe a substantial enhancement in prediction accuracy when using the individual components. This improvement is consistently evident across individual, pooled, and fixed effects regression models. To shed light on these findings, we computed the pooled correlation between returns and earnings for the entire sample, i.e.\ Corr($r_{i,t+1}, e^a_{i,t+1|t}$) = -0.206. The correlation indicates a (weak) negative relationship between returns and earnings. Consequently, the prediction errors of each component tend to offset each other, resulting in more accurate aggregated nowcasts (recall equation (\ref{eq:MSEcomponents})). The last two columns of Table \ref{tab:nowcasts} present the prediction results for these components. We observe that analyst earnings prediction errors appear to be more predictable than those of log returns.  We also report \cite{diebold1995comparing} test statistic p-values comparing each model against the RW and consensus benchmarks, pooling all the nowcasting errors across firms. Using one-sided test critical values we observe that our models outperform both the RW and consensus benchmarks, particularly when we use the component approach. While we cannot compare the $\hat pe_{i,t+1}$ component with the consensus, judging by the RW benchmark it is clear that the second component is the most important in terms of nowcasting gains. When we use individual MIDAS regressions the evidence is less compelling, underscoring the importance of using panel data models.\footnote{We also experimented with the forecast combination of MIDAS regressions used by \cite{ball2018automated} and found them to be inferior to the individual MIDAS ML regressions as well as the panel data models. We therefore refrain from reporting the details here. \label{footnote:forcomb}}

\begin{center}
	{\it Sparsity Patterns}
\end{center}

\begin{figure}[h]
	\centering
	\begin{subfigure}{0.32\textwidth} 
		\includegraphics[width=\textwidth]{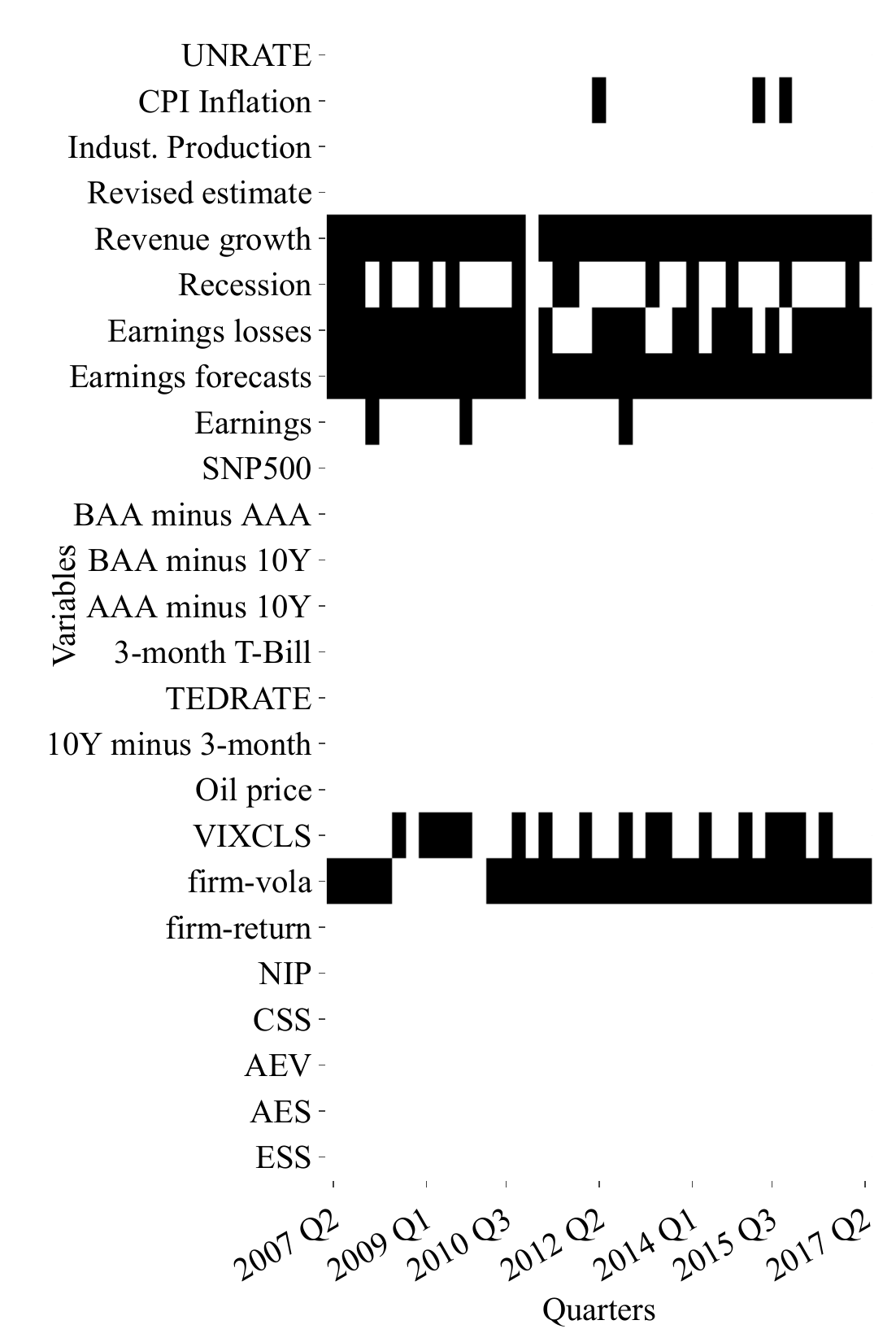}
		\caption{Best model for $\hat{pe}_{i,t+1}$.} 
	\end{subfigure}
	\begin{subfigure}{0.32\textwidth} 
		\includegraphics[width=\textwidth]{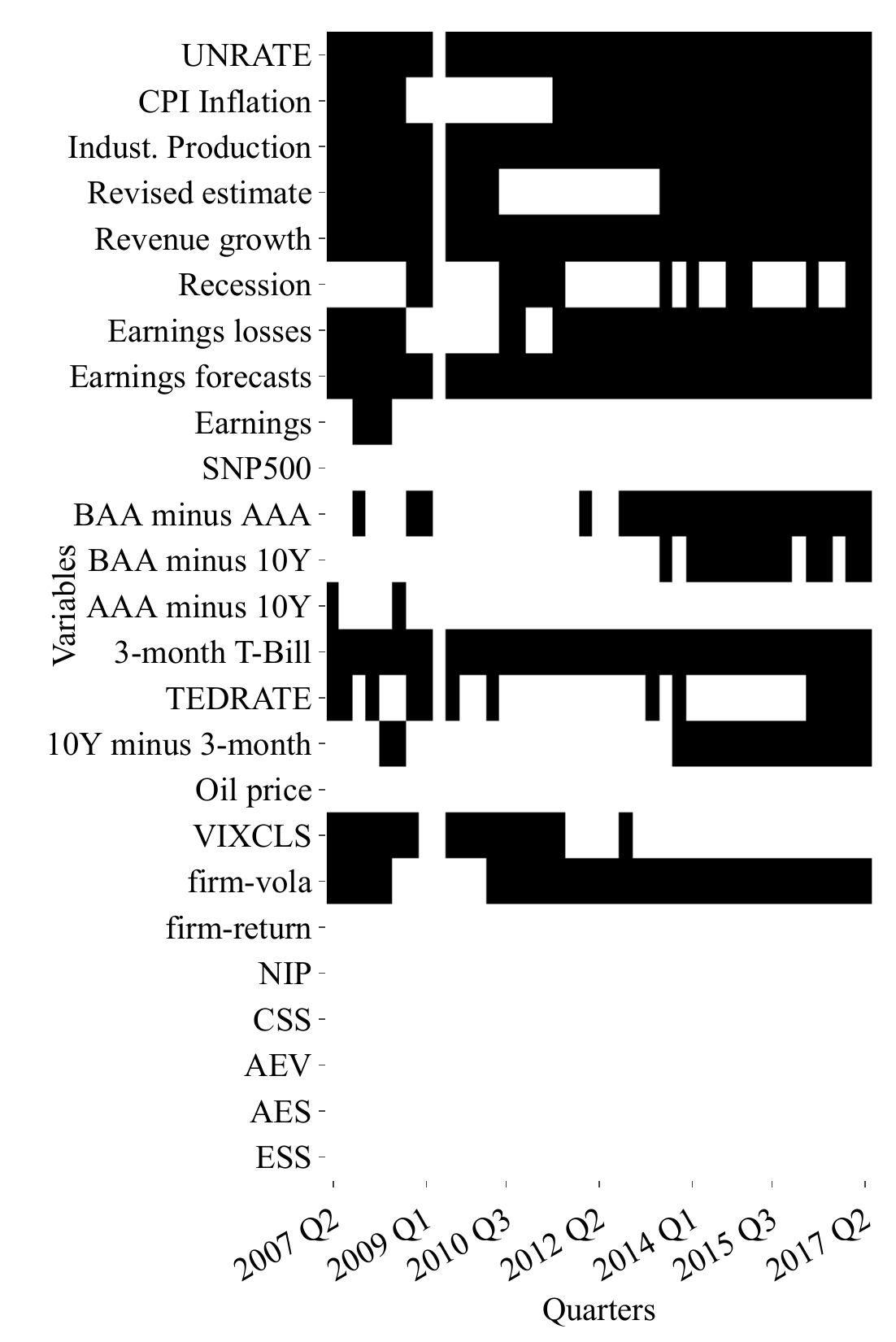}
		\caption{Best model for $\hat r_{i,t+1}$.} 
	\end{subfigure}
	\begin{subfigure}{0.32\textwidth} 
		\includegraphics[width=\textwidth]{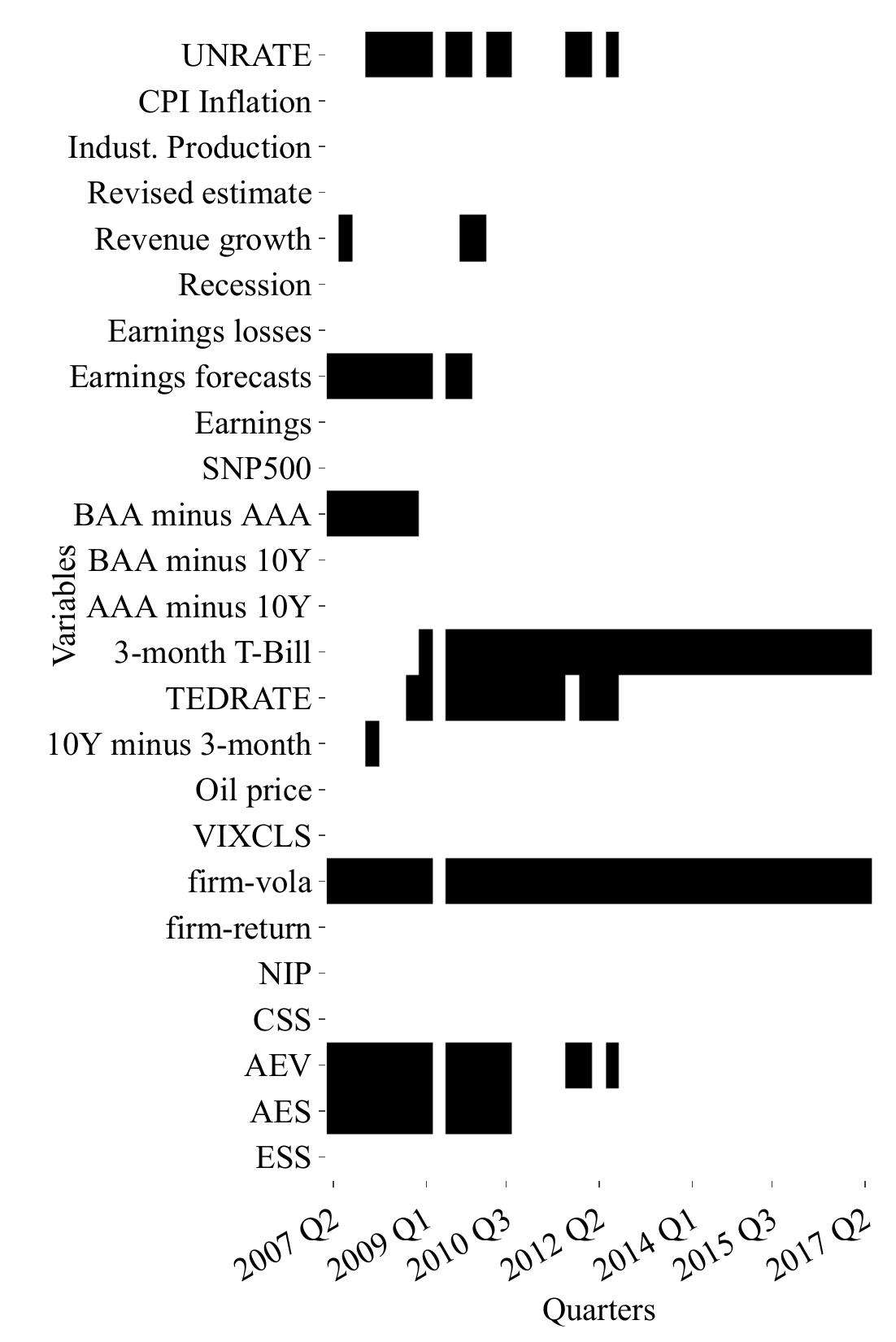}
		\caption{Best model for $\hat e^a_{i,t+1|t}$.} 
	\end{subfigure}
	\caption{Sparsity patterns.} 
	\label{fig:sparse_betas}
\end{figure}

Figure \ref{fig:sparse_betas} illustrates the sparsity patterns of selected covariates for the most effective methods in predicting either log P/E ratios (Panel a) or their components (Panels b and c). It is worth noting that the sparsity patterns differ significantly across the three panels. For instance, firm volatility is often chosen as a relevant covariate across all targets, albeit not consistently throughout the entire out-of-sample period. In the case of log P/E ratios, news series related to earnings are frequently selected, along with firm and market volatility series. Conversely, for log returns, a denser pattern of covariate selection is observed, distinct from the other two cases. Interestingly, none of the news-based firm series are chosen for this target. Regarding log analyst earnings forecast errors, macroeconomic series such as the unemployment rate, short-term rates, and TED rate are frequently selected. Moreover, unlike log P/E ratios and returns, news-based firm series occasionally appear in the selected covariates for this target. The fact that macroeconomic series are drivers for nowcasting the $e^a_{i,t+1|t}$ component is a confirmation of the findings reported in \cite{ball2018automated}, \cite{carabias2018real} and \cite{babii2022machine}.

\medskip

\begin{figure}[htp!]
	\centering
	\includegraphics[scale=0.65]{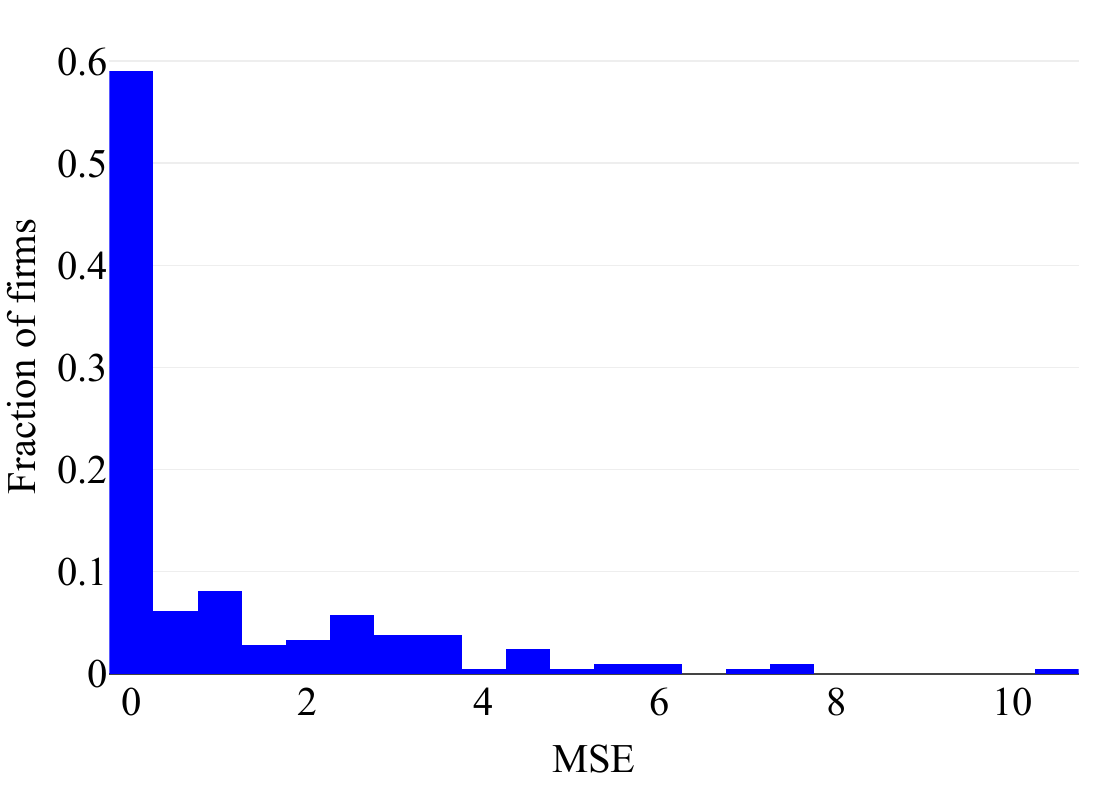}
	\caption{Histogram of mean squared errors.}  
	\label{fig:hist}
\end{figure}

Figure \ref{fig:hist} depicts the histogram of mean squared errors (MSEs) across firms. Notably, a substantial proportion of firms (approximately 60\%) exhibit low MSE values, indicating a high level of prediction accuracy. However, there are a few firms for which the MSEs are relatively larger, suggesting lower prediction performance for these specific cases.  The largest MSE is for Crown castle international corporation (CCI) which appears as a strong outlier. 

\medskip

Removing the single outlier firm has a dramatic impact on the nowcasting performance evaluation as shown in the lower panel of Table \ref{tab:nowcasts}. We now have very strong evidence that the panel regression models dominate analyst predictions. Again the component nowcasts are the best, but even the individual regression models do significantly better when the component specification is used.

\begin{center}
	{\it Daily Updates of Nowcasts}
\end{center}

\begin{figure}[h]
	\centering
	\includegraphics[scale=0.65]{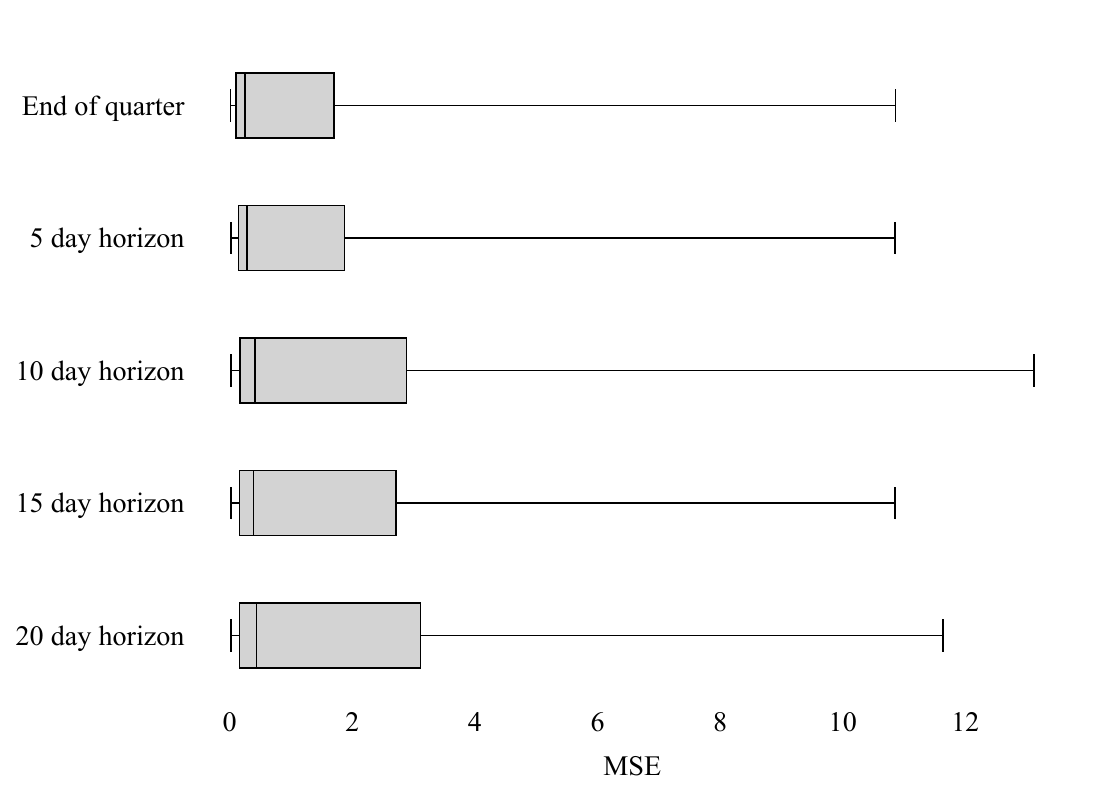}
	\caption{Distribution of MSEs of the best performing model in Table \ref{tab:nowcasts}. Models are re-estimated for each horizon. The best model based on Table \ref{tab:nowcasts} is reported.}  
	\label{fig:daily-updates}
\end{figure}

Our framework allows us to go beyond providing quarterly nowcasts and generate daily updates of earnings series. Leveraging the daily influx of information throughout the quarter, we continuously re-estimate our models and produce nowcast updates as soon as new data becomes available. In Figure \ref{fig:daily-updates}, we present the distribution of Mean Squared Errors (MSEs) across firms for five distinct nowcast horizons: 20-day, 15-day, 10-day, and 5-day ahead, as well as the end of the quarter. We report the best model based on Table \ref{tab:nowcasts}. Notably, as the horizons become shorter, both the median and upper quartile of MSEs decrease. Therefore, updating nowcasts with daily information appears to significantly enhance the prediction performance of log earnings ratios. The largest errors persist for the same firm, CCI.

\begin{center}
	{\it Grouped Fixed Effects based on \\
		Fama-French Industry Classification}
\end{center}

The sg-LASSO estimator we employ in our study is well-suited for incorporating grouped fixed effects. This approach involves grouping firm-specific intercepts based on either statistical procedures or economic reasoning, as outlined in \cite{bonhomme2015grouped}. In our analysis, we utilize the Fama French industry classification to form 10 distinct groups for grouping fixed effects. Rather than assuming a common fixed effect for all firms within a group, we apply a group penalty to the fixed effects of firms belonging to the same industry. This allows us to capture industry-specific heterogeneity while avoiding overfitting. 

\smallskip

We present the findings in Table \ref{tab:grouped_fes}, which highlight several key observations. Similar to previous analyses, our results suggest that predicting individual components of the log price-earnings ratio leads to more accurate aggregate nowcasts compared to a direct nowcast approach. 
Furthermore, we observe that the use of group fixed effects improves the accuracy of our nowcasts when forecasting individual components. This can be seen in column 2 of both Tables \ref{tab:nowcasts} and \ref{tab:grouped_fes}. Comparatively, when considering the best tuning parameter choice, grouped fixed effects outperform other panel models, including the pooled panel model. Therefore, our findings suggest that grouped fixed effects strike a better balance between capturing heterogeneity and pooled parameters, resulting in more accurate nowcast predictions. These results support the notion that incorporating group fixed effects enhances the overall performance of our forecasting model.

\begin{table}[ht]
	\centering
	\begin{tabular}{rrr}
		\hline
		\hline
		& $\hat pe_{i,t+1}$ & $\hat S$\\ 
		&\multicolumn{2}{c}{\textit{Group fixed effects}} \\ 
		CV & 0.862 & 0.789 \\ 
		BIC & 0.834 & 0.789 \\ 
		AIC & 0.842 & 0.791 \\ 
		AICc & 0.842 & 0.790 \\ 
		\hline
	\end{tabular}
	\caption{Nowcasting results. Column $\hat pe_{i,t+1}$ reports results for directly nowcasting the log P/E ratio and the column $\hat S$ reports the results of nowcasting and summing up the components. Results are reported relative to the {\it Consensus} nowcasts that appear in Table \ref{tab:nowcasts}. \label{tab:grouped_fes}}
\end{table}

\begin{figure}[h]
	\centering
	\includegraphics[scale=0.65]{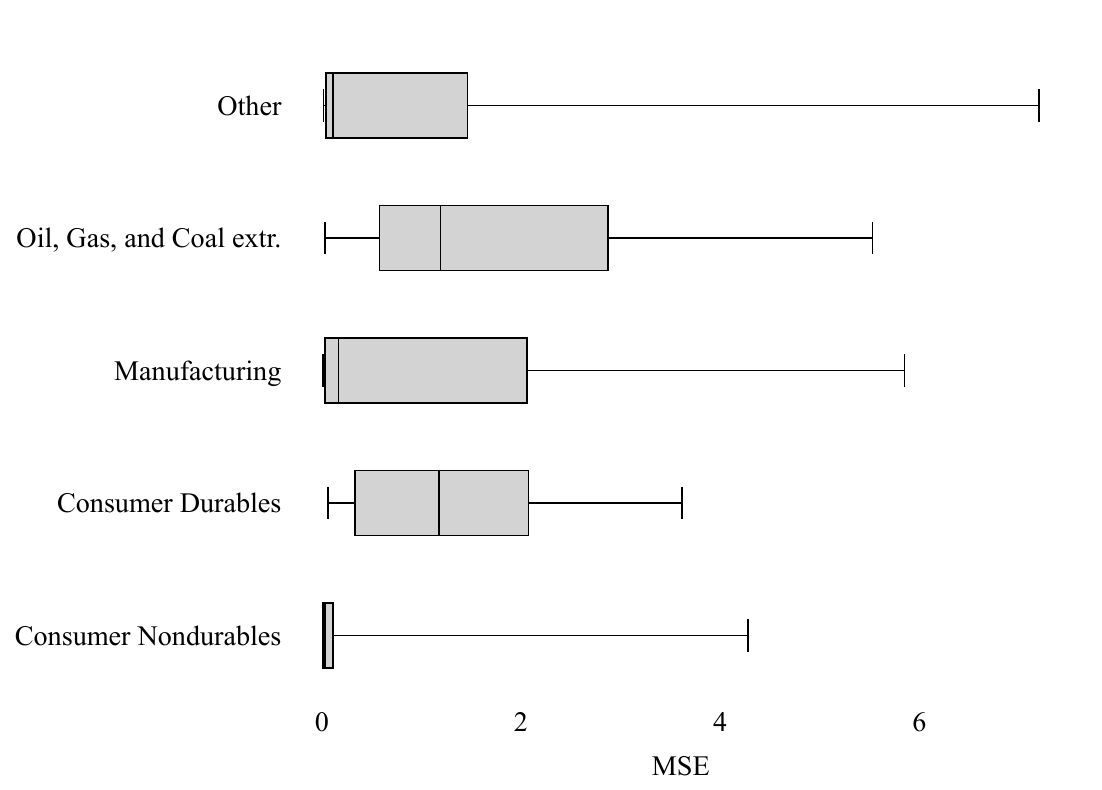}
	\caption{Distribution of MSEs for five industries based on Fama French classification. The reported resutls are based on the best model specification from Table \ref{tab:grouped_fes}.}  
	\label{fig:box}
\end{figure}

In Figure \ref{fig:box}, we present the distribution of (MSEs) across firms for five industries, based on the best model specification from Table \ref{tab:grouped_fes}. The industries we focus on are the ones with the highest number of firms in our sample. The results reveal variations in performance among different industries. Specifically, the firms categorized as {\it Consumer Durables} exhibit the lowest accuracy in terms of the median MSE, although the quartiles are comparatively lower compared to the other industries. On the other hand, the nowcasts for firms in the {\it Consumer Nondurables} and {\it Others} categories demonstrate the highest accuracy at the median. However, it is important to note that the largest errors occur within the firms classified as {\it Others}.

\begin{center}
	{\it Nowcasting with Missing Data — Parameter Imputation Method}
\end{center}

Next we address the challenge of missing earnings data, which can complicate the analysis. We examine the performance of parameter imputation methods in computing nowcasts, see, e.g, \cite{brown2023nowcasting}, even when earnings and/or earnings forecasts are missing for certain observations in the sample. We identify a subset of 117 firms for which at least one earnings observation is available in our out-of-sample period, and for which we have matched daily news data. To handle missing data, we match these firms with missing observations to firms in our main sample using the Fama French industry classification. We then utilize the parameter estimates obtained from the best group fixed effects model, as shown in Table \ref{tab:grouped_fes}, to compute the nowcasts of log earnings ratios, either directly or based on its components. The results of this analysis appear in Table \ref{tab:param_imput}.

\medskip

Firstly, the results obtained through parameter imputation support the conclusion that nowcasting the components of the log earnings ratio yields higher quality predictions. This indicates that incorporating the individual components of the ratio improves the accuracy of the nowcasts. Secondly, the panel models with the parameter imputation method outperform the analyst consensus nowcasts in terms of prediction accuracy. This suggests that employing machine learning panel data models along with parameter imputation could be a straightforward yet effective approach in situations where earnings data is not available. Overall, these findings highlight the potential benefits of leveraging machine learning techniques and imputation methods for improving nowcasting accuracy, particularly in cases where earnings data may be missing.

\begin{table}[ht]
	\centering
	\begin{tabular}{rrr}
		\hline
		\hline
		& $\hat pe_{i,t+1}$ & $\hat S$\\ \ 
		Consensus& 1.605  & \\ 
		CV & 0.883 & 0.753 \\ 
		BIC & 0.877 & 0.756 \\ 
		AIC & 0.883 & 0.754 \\ 
		AICc & 0.883 & 0.753 \\ 
		\hline
	\end{tabular}
		\caption{Nowcasting results — parameter imputation method. Column $\hat pe_{i,t+1}$ reports results for directly nowcasting the log P/E ratio and the column $\hat S$ reports the results of nowcasting and summing up the components. Row {\it Consensus} for the median consensus nowcast. Panels {\it Individual},  {\it Pooled} and {\it Fixed effects} report results for different panel data models relative to the consensus MSE. \label{tab:param_imput}}
\end{table}

\section{Conclusions \label{sec:conclusion}}

This paper uses a new class of high-dimensional panel data nowcasting models with dictionaries and sg-LASSO regularization which is an attractive choice for the predictive panel data regressions, where the low- and/or the high-frequency lags define a clear group structure. 
Our empirical results showcase the advantages of using regularized panel data regressions for nowcasting corporate earnings either directly or using a decomposition which separates stock market return predictions and analyst assessments of a firm's performance. While nowcasting earnings is a leading example of applying panel data MIDAS machine learning regressions, one can think of many other applications of interest in finance. Beyond earnings, analysts are also interested in sales, dividends, etc. Our analysis can also be useful for other areas of interest, such as regional and international panel data settings.

\clearpage 
\bibliographystyle{econometrica}
\bibliography{midas_ml}

\newpage

\setcounter{page}{1}
\setcounter{section}{0}
\setcounter{equation}{0}
\setcounter{table}{0}
\setcounter{figure}{0}
\renewcommand{\theequation}{OA.\arabic{equation}}
\renewcommand\thetable{OA.\arabic{table}}
\renewcommand\thefigure{OA.\arabic{figure}}
\renewcommand\thesection{OA.\arabic{section}}
\renewcommand\thepage{Online Appendix - \arabic{page}}
\renewcommand\thetheorem{OA.\arabic{theorem}}

\begin{center}
	{\LARGE\textbf{ONLINE APPENDIX}}	
\end{center}
\bigskip

\section{Additional simulation results \label{app:mcs}}

\begin{table}[!htbp]
	\centering
	\scriptsize{
		\setlength{\tabcolsep}{4.5pt}
		\begin{tabular}{rcrrrrrr r  rrrrrr}
			\rule{0pt}{4ex} 
			&&\multicolumn{6}{c}{\underline{Pooled panel data}} &&\multicolumn{6}{c}{\underline{Fixed effects}}\\
			\rule{0pt}{4ex} 
			& N/T & $\gamma$ = 0 & 0.2 & 0.4 & 0.6 & 0.8 & 1  &  & $\gamma$ =  0 & 0.2 & 0.4 & 0.6 & 0.8 & 1\\
			\hline
			\rule{0pt}{3ex} 
			&&\multicolumn{13}{c}{\underline{Cross-validation}} \\ 
			\rule{0pt}{3ex} 
			&25/50  & 1.194 & 1.191 & 1.192 & 1.200 & 1.212 & 1.207  & & 1.202 & 1.200 & 1.198 & 1.199 & 1.205 & 1.210 \\
			sg-LASSO &75/50& 1.160 & 1.162 & 1.159 & 1.157 & 1.165 & 1.161 & & 1.171 & 1.171 & 1.170 & 1.171 & 1.173 & 1.174 \\
			&25/100 & 1.175 & 1.174 & 1.175 & 1.172 & 1.168 & 1.183 & & 1.166 & 1.165 & 1.165 & 1.164 & 1.166 & 1.171\\\rule{0pt}{3ex} 
			&25/50 & 1.334 & 1.216 & 1.213 & 1.213 & 1.214 & 1.214 & & 1.370 & 1.248 & 1.245 & 1.245 & 1.246 & 1.245 \\
			elnet-U &75/50 & 1.209 & 1.160 & 1.158 & 1.158 & 1.158 & 1.158 & & 1.236 & 1.186 & 1.184 & 1.183 & 1.183 & 1.183 \\
			&25/100 & 1.242 & 1.172 & 1.172 & 1.172 & 1.172 & 1.173 & & 1.255 & 1.184 & 1.184 & 1.184 & 1.184 & 1.185 \\
			\rule{0pt}{3ex}   
			&&\multicolumn{13}{c}{\underline{BIC}} \\ 
			\rule{0pt}{3ex} 
			&25/50  & 1.272 & 1.270 & 1.273 & 1.289 & 1.315 & 1.357 & & 1.310 & 1.304 & 1.317 & 1.346 & 1.410 & 1.471 \\
			sg-LASSO &75/50 & 1.177 & 1.175 & 1.177 & 1.180 & 1.180 & 1.199  & & 1.202 & 1.202 & 1.208 & 1.209 & 1.222 & 1.282 \\
			&25/100 & 1.207 & 1.202 & 1.203 & 1.205 & 1.223 & 1.260 & & 1.213 & 1.213 & 1.220 & 1.219 & 1.247 & 1.307   \\\rule{0pt}{3ex}  	
			&25/50  & 1.524 & 1.411 & 1.388 & 1.385 & 1.384 & 1.385 & & 1.537 & 1.575 & 1.595 & 1.611 & 1.626 & 1.639 \\
			elnet-U &75/50 & 1.236 & 1.226 & 1.216 & 1.213 & 1.211 & 1.211 & & 1.259 & 1.274 & 1.276 & 1.278 & 1.279 & 1.280 \\
			&25/100  & 1.298 & 1.255 & 1.247 & 1.248 & 1.248 & 1.249  & & 1.313 & 1.318 & 1.318 & 1.318 & 1.320 & 1.320 \\
			\rule{0pt}{3ex}    
			&&\multicolumn{13}{c}{\underline{AIC}} \\ 
			\rule{0pt}{3ex} 
			&25/50  & 1.252 & 1.247 & 1.242 & 1.234 & 1.245 & 1.265 & & 1.288 & 1.282 & 1.283 & 1.287 & 1.284 & 1.303 \\
			sg-LASSO &75/50 & 1.167 & 1.165 & 1.163 & 1.160 & 1.166 & 1.166 & & 1.199 & 1.196 & 1.197 & 1.198 & 1.192 & 1.205 \\ 
			&25/100 & 1.193 & 1.190 & 1.191 & 1.196 & 1.187 & 1.200 & & 1.200 & 1.196 & 1.196 & 1.202 & 1.209 & 1.208 \\  \\\rule{0pt}{3ex}  	
			&25/50  & 1.524 & 1.282 & 1.274 & 1.273 & 1.273 & 1.274 & & 1.537 & 1.388 & 1.381 & 1.380 & 1.378 & 1.378 \\ 
			elnet-U &75/50 & 1.236 & 1.179 & 1.174 & 1.173 & 1.172 & 1.172 & & 1.259 & 1.222 & 1.223 & 1.222 & 1.222 & 1.222 \\
			&25/100  & 1.298 & 1.213 & 1.215 & 1.216 & 1.215 & 1.215 & & 1.313 & 1.255 & 1.243 & 1.240 & 1.238 & 1.237 \\
			\rule{0pt}{3ex} 
			&&\multicolumn{13}{c}{\underline{AICc}} \\ 
			\rule{0pt}{3ex} 
			&25/50  & 1.255 & 1.248 & 1.244 & 1.237 & 1.246 & 1.270 & & 1.291 & 1.284 & 1.285 & 1.290 & 1.291 & 1.307 \\ 
			sg-LASSO &75/50  & 1.168 & 1.165 & 1.164 & 1.161 & 1.166 & 1.167 & & 1.200 & 1.197 & 1.197 & 1.199 & 1.193 & 1.205 \\
			&25/100 & 1.193 & 1.190 & 1.191 & 1.196 & 1.188 & 1.201 & & 1.200 & 1.196 & 1.196 & 1.202 & 1.210 & 1.210  \\\rule{0pt}{3ex}  	
			&25/50 & 1.524 & 1.286 & 1.279 & 1.280 & 1.208 & 1.281 & & 1.537 & 1.410 & 1.396 & 1.390 & 1.388 & 1.387 \\
			elnet-U &75/50  & 1.236 & 1.179 & 1.174 & 1.173 & 1.172 & 1.172 & & 1.259 & 1.227 & 1.226 & 1.226 & 1.225 & 1.225 \\
			&25/100  & 1.298 & 1.217 & 1.219 & 1.218 & 1.218 & 1.218  & & 1.313 & 1.257 & 1.244 & 1.240 & 1.238 & 1.237 \\
			\hline
		\end{tabular}
	}
	\caption{\small The table reports the MSFE for nowcasting accuracy for pooled and fixed effects estimators for the baseline DGP for the sg-LASSO-MIDAS (rows sg-LASSO) and elastic net UMIDAS (rows elnet-U). We vary the cross-sectional dimension $N\in\{25, 75\}$ and time series dimension $T\in\{50, 100\}$.  We report results for 5-fold cross-validation, BIC, AIC, AICc information criteria $\lambda$ tuning parameter calculation methods and for a grid of $\gamma \in \{0, 0.2, \dots, 1\}$ tuning parameter. \label{app:main_mcres1}} 
\end{table}

\begin{table}[!htbp]
	\centering
	\scriptsize{
		\setlength{\tabcolsep}{4.5pt}
		\begin{tabular}{rcrrrrrr r  rrrrrr}
			\rule{0pt}{4ex} 
			&&\multicolumn{6}{c}{\underline{Pooled panel data}} &&\multicolumn{6}{c}{\underline{Fixed effects}}\\
			\rule{0pt}{4ex} 
			& N/T & $\gamma$ = 0 & 0.2 & 0.4 & 0.6 & 0.8 & 1  &  & $\gamma$ =  0 & 0.2 & 0.4 & 0.6 & 0.8 & 1\\
			\hline
			\rule{0pt}{3ex} 
			&&\multicolumn{13}{c}{\underline{Cross-validation}} \\ 
			\rule{0pt}{3ex} 
			&25/50  & 1.289 & 1.288 & 1.288 & 1.282 & 1.280 & 1.302 & & 1.282 & 1.279 & 1.278 & 1.280 & 1.282 & 1.291 \\
			sg-LASSO &75/50 & 1.247 & 1.245 & 1.247 & 1.253 & 1.246 & 1.255 & & 1.256 & 1.256 & 1.256 & 1.256 & 1.257 & 1.261 \\
			&25/100 & 1.250 & 1.248 & 1.248 & 1.254 & 1.264 & 1.257 & & 1.249 & 1.248 & 1.248 & 1.249 & 1.252 & 1.255   \\\rule{0pt}{3ex} 
			&25/50 & 1.452 & 1.302 & 1.299 & 1.300 & 1.300 & 1.301 & & 1.487 & 1.334 & 1.331 & 1.331 & 1.330 & 1.329 \\ 
			elnet-U &75/50 & 1.304 & 1.244 & 1.243 & 1.243 & 1.243 & 1.243 & & 1.333 & 1.271 & 1.270 & 1.270 & 1.270 & 1.270 \\ 
			&25/100& 1.342 & 1.258 & 1.257 & 1.256 & 1.256 & 1.256 & & 1.359 & 1.273 & 1.271 & 1.271 & 1.271 & 1.271 \\ 
			\rule{0pt}{3ex}   
			&&\multicolumn{13}{c}{\underline{BIC}} \\ 
			\rule{0pt}{3ex} 
			&25/50  & 1.395 & 1.389 & 1.395 & 1.401 & 1.444 & 1.517 & & 1.437 & 1.437 & 1.450 & 1.468 & 1.553 & 1.657 \\
			sg-LASSO &75/50 & 1.275 & 1.274 & 1.278 & 1.278 & 1.281 & 1.310 & & 1.306 & 1.309 & 1.319 & 1.321 & 1.317 & 1.370 \\ 
			&25/100 & 1.296 & 1.293 & 1.297 & 1.306 & 1.308 & 1.364 & & 1.311 & 1.310 & 1.319 & 1.325 & 1.342 & 1.431  \\\rule{0pt}{3ex}  	
			&25/50  & 1.652 & 1.582 & 1.570 & 1.570 & 1.572 & 1.575 & & 1.694 & 1.754 & 1.781 & 1.803 & 1.813 & 1.820 \\
			elnet-U &75/50 & 1.337 & 1.318 & 1.317 & 1.318 & 1.319 & 1.319 & & 1.367 & 1.401 & 1.393 & 1.393 & 1.394 & 1.396 \\
			&25/100  & 1.389 & 1.387 & 1.373 & 1.369 & 1.368 & 1.367 & & 1.404 & 1.430 & 1.427 & 1.432 & 1.434 & 1.436 \\
			\rule{0pt}{3ex}    
			&&\multicolumn{13}{c}{\underline{AIC}} \\ 
			\rule{0pt}{3ex} 
			&25/50  & 1.354 & 1.345 & 1.342 & 1.351 & 1.370 & 1.372 & & 1.397 & 1.396 & 1.393 & 1.389 & 1.407 & 1.450 \\ 
			sg-LASSO &75/50 & 1.261 & 1.259 & 1.259 & 1.264 & 1.271 & 1.266 & & 1.292 & 1.293 & 1.294 & 1.292 & 1.301 & 1.309 \\
			&25/100 & 1.282 & 1.280 & 1.279 & 1.272 & 1.279 & 1.290 & & 1.305 & 1.301 & 1.302 & 1.300 & 1.294 & 1.320   \\\rule{0pt}{3ex}  	
			&25/50  & 1.652 & 1.435 & 1.419 & 1.414 & 1.412 & 1.411 & & 1.694 & 1.489 & 1.480 & 1.479 & 1.478 & 1.478 \\
			elnet-U &75/50 & 1.337 & 1.293 & 1.287 & 1.285 & 1.284 & 1.283 & & 1.367 & 1.327 & 1.319 & 1.317 & 1.316 & 1.316 \\
			&25/100  & 1.389 & 1.304 & 1.298 & 1.298 & 1.298 & 1.298 & & 1.404 & 1.342 & 1.342 & 1.342 & 1.343 & 1.343 \\
			\rule{0pt}{3ex} 
			&&\multicolumn{13}{c}{\underline{AICc}} \\ 
			\rule{0pt}{3ex} 
			&25/50  & 1.357 & 1.348 & 1.344 & 1.352 & 1.372 & 1.377 & & 1.402 & 1.402 & 1.398 & 1.393 & 1.409 & 1.459 \\
			sg-LASSO &75/50  & 1.261 & 1.259 & 1.259 & 1.264 & 1.271 & 1.266 & & 1.293 & 1.295 & 1.295 & 1.293 & 1.301 & 1.312 \\
			&25/100 & 1.282 & 1.280 & 1.279 & 1.273 & 1.279 & 1.292 & & 1.306 & 1.302 & 1.303 & 1.301 & 1.295 & 1.321 \\\rule{0pt}{3ex}  	
			&25/50 & 1.652 & 1.436 & 1.420 & 1.415 & 1.414 & 1.412 & & 1.694 & 1.503 & 1.495 & 1.496 & 1.496 & 1.495 \\
			elnet-U &75/50 & 1.337 & 1.294 & 1.287 & 1.285 & 1.284 & 1.283 & & 1.367 & 1.328 & 1.320 & 1.318 & 1.317 & 1.316 \\ 
			&25/100  & 1.389 & 1.305 & 1.300 & 1.300 & 1.300 & 1.300 & & 1.404 & 1.348 & 1.348 & 1.348 & 1.348 & 1.348 \\
			\hline
		\end{tabular}
	}
	\caption{\small The table reports the MSFE for nowcasting accuracy for pooled and fixed effects estimators for the student-$t(5)$ DGP for the sg-LASSO-MIDAS (rows sg-LASSO) and elastic net UMIDAS (rows elnet-U). We vary the cross-sectional dimension $N\in\{25, 75\}$ and time series dimension $T\in\{50, 100\}$.  We report results for 5-fold cross-validation, BIC, AIC, AICc information criteria $\lambda$ tuning parameter calculation methods and for a grid of $\gamma \in \{0, 0.2, \dots, 1\}$ tuning parameter. \label{app:main_mcres2}} 
\end{table}

\begin{table}[!htbp]
	\centering
	\scriptsize{
		\setlength{\tabcolsep}{4.5pt}
		\begin{tabular}{rcrrrrrr r  rrrrrr}
			\rule{0pt}{4ex} 
			&&\multicolumn{6}{c}{\underline{Pooled panel data}} &&\multicolumn{6}{c}{\underline{Fixed effects}}\\
			\rule{0pt}{4ex} 
			& N/T & $\gamma$ = 0 & 0.2 & 0.4 & 0.6 & 0.8 & 1  &  & $\gamma$ =  0 & 0.2 & 0.4 & 0.6 & 0.8 & 1\\
			\hline
			\rule{0pt}{3ex} 
			&&\multicolumn{13}{c}{\underline{Cross-validation}} \\ 
			\rule{0pt}{3ex} 
			&25/50  & 1.208 & 1.207 & 1.204 & 1.207 & 1.226 & 1.249 &  &  1.217 & 1.214 & 1.215 & 1.218 & 1.225 & 1.242\\
			sg-LASSO &75/50 & 1.168 & 1.168 & 1.168 & 1.160 & 1.161 & 1.178 &  &   1.171 & 1.170 & 1.170 & 1.173 & 1.180 &  1.187\\
			&25/100 & 1.188 & 1.185 & 1.187 & 1.194 & 1.186 & 1.199 &  &  1.173 &  1.172 & 1.172 & 1.174 & 1.176 & 1.181  \\\rule{0pt}{3ex} 
			&25/50 & 1.291 & 1.263 & 1.257 & 1.255 & 1.255 & 1.255 &  &   1.292 & 1.290 & 1.285 & 1.283 & 1.283 & 1.282\\
			elnet-U &75/50 & 1.259 & 1.170 & 1.167 & 1.166 & 1.165 & 1.165 &  &   1.273 & 1.204 & 1.199 & 1.198 & 1.197 & 1.197\\
			&25/100 & 1.258 & 1.196 & 1.190 & 1.189 & 1.189 & 1.188 &  &  1.999 & 1.199 & 1.196 & 1.194 & 1.194 & 1.193\\
			\rule{0pt}{3ex}   
			&&\multicolumn{13}{c}{\underline{BIC}} \\ 
			\rule{0pt}{3ex} 
			&25/50  & 1.273 & 1.274 & 1.278 & 1.295 & 1.336 & 1.408 & & 1.358 & 1.344 & 1.353 & 1.374 & 1.443 & 1.505 \\
			sg-LASSO &75/50 & 1.176 & 1.175 & 1.177 & 1.184 & 1.200 & 1.234 & & 1.215 & 1.213 & 1.218 & 1.218 & 1.237 & 1.305 \\
			&25/100 & 1.226 & 1.218 & 1.217 & 1.214 & 1.227 & 1.273 & & 1.232 & 1.229 & 1.236 & 1.240 & 1.267 & 1.332   \\\rule{0pt}{3ex}  	
			&25/50  & 1.524 & 1.428 & 1.409 & 1.412 & 1.413 & 1.417 & & 1.689 & 1.675 & 1.662 & 1.670 & 1.682 & 1.689 \\
			elnet-U &75/50 & 1.359 & 1.229 & 1.215 & 1.211 & 1.209 & 1.208 & & 1.306 & 1.298 & 1.299 & 1.301 & 1.302 & 1.303 \\
			&25/100   & 1.304 & 1.297 & 1.209 & 1.209 & 1.289 & 1.290 & & 1.414 & 1.372 & 1.349 & 1.343 & 1.342 & 1.342 \\ 
			\rule{0pt}{3ex}    
			&&\multicolumn{13}{c}{\underline{AIC}} \\ 
			\rule{0pt}{3ex} 
			&25/50  & 1.263 & 1.259 & 1.264 & 1.272 & 1.264 & 1.285 & & 1.307 & 1.300 & 1.300 & 1.315 & 1.343 & 1.380 \\
			sg-LASSO &75/50 & 1.175 & 1.172 & 1.174 & 1.176 & 1.166 & 1.185 & & 1.250& 1.246 & 1.245& 1.243 & 1.250 & 1.261 \\
			&25/100 & 1.195 & 1.191 & 1.191 & 1.200 & 1.217 & 1.216 & & 1.210 & 1.205 & 1.202 & 1.203 & 1.222 & 1.251   \\\rule{0pt}{3ex}  	
			&25/50  & 1.527 & 1.381 & 1.361 & 1.354 & 1.351 & 1.350 & & 1.559 & 1.458 & 1.421 & 1.411 & 1.406 & 1.404 \\
			elnet-U &75/50 & 1.359 & 1.201 & 1.200 & 1.199 & 1.199 & 1.198 & & 1.401 & 1.400 & 1.385 & 1.384 & 1.384 & 1.392 \\
			&25/100  & 1.301 & 1.252 & 1.239 & 1.235 & 1.233 & 1.232 & & 1.314 & 1.255 & 1.241 & 1.238 & 1.236 & 1.235 \\
			\rule{0pt}{3ex} 
			&&\multicolumn{13}{c}{\underline{AICc}} \\ 
			\rule{0pt}{3ex} 
			&25/50  & 1.264 & 1.260 & 1.264 & 1.274 & 1.267 & 1.286  & & 1.312 & 1.301 & 1.302 & 1.316 & 1.347 & 1.390 \\ 
			sg-LASSO &75/50 & 1.175 & 1.172 & 1.174 & 1.176 & 1.167 & 1.185 & & 1.209 & 1.205 & 1.205 & 1.212 & 1.224 & 1.228 \\ 
			&25/100 & 1.195 & 1.192 & 1.192 & 1.200 & 1.218 & 1.218  & & 1.211 & 1.206 & 1.204 & 1.203 & 1.222 & 1.263  \\\rule{0pt}{3ex}  	
			&25/50 & 1.524 & 1.390 & 1.364 & 1.358 & 1.354 & 1.353 & & 1.559 & 1.460 & 1.422 & 1.413 & 1.408 & 1.405 \\
			elnet-U &75/50 & 1.359 & 1.203 & 1.202 & 1.201 & 1.200 & 1.200 & & 1.306 & 1.271 & 1.255 & 1.250 & 1.248 & 1.247 \\ 
			&25/100  & 1.302 & 1.252 & 1.239 & 1.235 & 1.233 & 1.232 & & 1.314 & 1.256 & 1.242 & 1.239 & 1.238 & 1.238 \\ 
			\hline
		\end{tabular}
	}
	\caption{\small The table reports the MSFE for nowcasting accuracy for pooled and fixed effects estimators for the large-dimensional DGP for the sg-LASSO-MIDAS (rows sg-LASSO) and elastic net UMIDAS (rows elnet-U). We vary the cross-sectional dimension $N\in\{25, 75\}$ and time series dimension $T\in\{50, 100\}$.  We report results for 5-fold cross-validation, BIC, AIC, AICc information criteria $\lambda$ tuning parameter calculation methods and for a grid of $\gamma \in \{0, 0.2, \dots, 1\}$ tuning parameter. \label{app:main_mcres3}} 
\end{table}

\clearpage

\section{Data description \label{appsec:data-description}}

\subsection{Firm-level data} 

The full list of firm-level data is provided in Table \ref{tab:firm_data}. 
We also add two daily firm-specific stock market predictor variables: stock returns and a realized variance measure, which is defined as the rolling sample variance over the previous 60 days (i.e.\ 60-day historical volatility).

\subsubsection{Firm sample selection}

We select a sample of firms based on data availability. First, we remove all firms from I/B/E/S which have missing values in earnings time series. Next, we retain firms that we are able to match with CRSP dataset. Finally, we keep firms that we can match with the RavenPack dataset.

\subsubsection{Firm-specific text data}

We create a link table of RavenPack ID and PERMNO identifiers which enables us to merge I/B/E/S and CRSP data with firm-specific textual analysis generated data from RavenPack. The latter is a rich dataset that contains intra-daily news information about firms. There are several editions of the dataset; in our analysis, we use the Dow Jones (DJ) and Press Release (PR) editions. The former contains relevant information from Dow Jones Newswires, regional editions of the Wall Street Journal, Barron's and MarketWatch. The PR edition contains news data, obtained from various press releases and regulatory disclosures, on a daily basis from a variety of newswires and press release distribution networks, including exclusive content from PRNewswire, Canadian News Wire, Regulatory News Service, and others. The DJ edition sample starts at $1^{st}$ of January, 2000, and PR edition data starts at $17^{th}$ of January, 2004. 

\smallskip

We construct our news-based firm-level covariates by filtering only highly relevant news stories. More precisely, for each firm and each day, we filter out news that has the \emph{Relevance Score} (REL) larger or equal to 75, as is suggested by the RavenPack News Analytics guide and used by practitioners, see for example \cite{kolanovic2017big}. REL is a score between 0 and 100 which indicates how strongly a news story is linked with a particular firm. A score of zero means that the entity is vaguely mentioned in the news story, while 100 means the opposite. A score of 75 is regarded as a significantly relevant news story. After applying the REL filter, we apply a novelty of the news filter by using the \emph{Event Novelty Score} (ENS); we keep data entries that have a score of 100. Like REL, ENS is a score between 0 and 100. It indicates the novelty of a news story within a 24-hour time window. A score of 100 means that a news story was not already covered by earlier announced news, while subsequently published news story score on a related event is discounted, and therefore its scores are less than 100. Therefore, with this filter, we consider only novel news stories. We focus on {\it five sentiment indices} that are available in both DJ and PR editions. They are: 

\paragraph{\bf Event Sentiment Score} (ESS), for a given firm, represents the strength of the news measured using surveys of financial expert ratings for firm-specific events. The score value ranges between 0 and 100 - values above (below) 50 classify the news as being positive (negative), 50 being neutral. 

\paragraph{\bf Aggregate Event Sentiment} (AES) represents the ratio of positive events reported on a firm compared to the total count of events measured over a rolling 91-day window in a particular news edition (DJ or PR). An event with ESS $>$ 50 is counted as a positive entry while ESS $<$ 50 as negative. Neutral news (ESS = 50) and news that does not receive an ESS score does not enter into the AES computation. As ESS, the score values are between 0 and 100. 

\paragraph{\bf Aggregate Event Volume} (AEV) represents the count of events for a firm over the last 91 days within a certain edition. As in AES case, news that receives a non-neutral ESS score is counted and therefore accumulates positive and negative news. 

\paragraph{\bf Composite Sentiment Score} (CSS) represents the news sentiment of a given news story by combining various sentiment analysis techniques. The direction of the score is determined by looking at emotionally charged words and phrases and by matching stories typically rated by experts as having short-term positive or negative share price impact. The strength of the scores is determined by intra-day price reactions modeled empirically using tick data from approximately 100 large-cap stocks. As for ESS and AES, the score takes values between 0 and 100, 50 being the neutral.  

\paragraph{\bf News Impact Projections} (NIP) represents the degree of impact a news flash has on the market over the following two-hour period. The algorithm produces scores to accurately predict a relative volatility - defined as scaled volatility by the average of volatilities of large-cap firms used in the test set - of each stock price measured within two hours following the news. Tick data is used to train the algorithm and produce scores, which take values between 0 and 100, 50 representing zero impact news.

\smallskip

For each firm and each day with firm-specific news, we compute the average value of the specific sentiment score. In this way, we aggregate across editions and groups, where the later is defined as a collection of related news. We then map the indices that take values between 0 and 100 onto $[-1,1]$. Specifically, let \(x_i \in \{\text{ESS}, \text{AES}, \text{CSS}, \text{NIP}\}\) be the average score value for a particular day and firm. We map $x_i \mapsto \bar x_i\in[-1,1]$ by computing
$\bar x_i$ = $(x_i - 50)/50.$



\clearpage

\begin{table}[!htbp]
	\centering
	{\scriptsize
		\begin{tabular}{r |l ccc}
			& id & Frequency & Source  & T-code \\ 
			\hline
			\hline
			\multicolumn{5}{c}{Panel A. } \\
			- & Price/Earnings ratio & quarterly & CRSP \& I/B/E/S & 1 \\
			- &Price/Earnings ratio consensus forecasts & quarterly & CRSP \& I/B/E/S & 1 \\
			\multicolumn{5}{c}{Panel B. } \\ 
			1 & Stock returns & daily & CRSP & 1 \\ 
			2 & Realized variance measure & daily & CRSP/computations & 1 \\ 
			\multicolumn{5}{c}{Panel C. } \\ 
			1 & Event Sentiment Score (ESS) & daily & RavenPack & 1 \\
			2 & Aggregate Event Sentiment (AES) & daily & RavenPack & 1 \\
			3 & Aggregate Event Volume (AEV) & daily & RavenPack & 1 \\
			4 & Composite Sentiment Score (CSS) & daily & RavenPack & 1 \\
			5 & News Impact Projections (NIP) & daily & RavenPack & 1 \\
			\hline 
		\end{tabular}
	}
	\caption{\footnotesize Firm-level data description table -- The \textit{id} column gives mnemonics according to data source, which is given in the second column \textit{Source}. The column \textit{frequency} states the sampling frequency of the variable. The column \textit{T-code} denotes the data transformation applied to a time-series, which are: (1) not transformed, (2) $\Delta x_t$, (3) $\Delta^2 x_t$, (4) log($x_t$), (5) $\Delta$ log ($x_t$), (6) $\Delta^2$ log ($x_t$). Panel A. describes earnings data, panel B. describes quarterly firm-level accouting data, panel C. daily firm-level stock market data and panel D. daily firm-level sentiment data series. \label{tab:firm_data}}
\end{table}

\begin{table}[!htbp]
	\centering
	{\scriptsize
		\begin{tabular}{r |l ccc}
			& id & Frequency & Source  & T-code \\ 
			\hline
			\hline
			\multicolumn{5}{c}{Panel A. } \\
			1 & Industrial Production Index & monthly & FRED-MD & 5 \\
			2 & CPI Inflation & monthly & FRED-MD & 6 \\
			\multicolumn{5}{c}{Panel B. } \\
			1 & Crude Oil Prices & daily & FRED & 6 \\ 
			2 & S\&P 500 & daily & CRSP & 5 \\ 
			3 & VIX Volatility Index & daily & FRED & 1 \\ 
			4 & Moodys Aaa - 10-Year Treasury & daily & FRED & 1 \\ 
			5 & Moodys Baa - 10-Year Treasury & daily & FRED & 1 \\ 
			6 & Moodys Baa - Aaa Corporate Bond & daily & FRED & 1 \\ 
			7 & 10-Year Treasury - 3-Month Treasury & daily & FRED & 1 \\ 
			8 & 3-Month Treasury - Effective Federal funds rate & daily & FRED & 1 \\ 
			9 & TED rate & daily & FRED & 1 \\ 
			\multicolumn{5}{c}{Panel C. } \\ 
			1 & Earnings & monthly & \cite{bybee2019structure} & 1 \\
			2 & Earnings forecasts & monthly & \cite{bybee2019structure} & 1 \\
			3 & Earnings losses & monthly & \cite{bybee2019structure} & 1 \\
			4 & Recession & monthly & \cite{bybee2019structure} & 1 \\
			5 & Revenue growth & monthly & \cite{bybee2019structure}& 1 \\
			6 & Revised estimate & monthly & \cite{bybee2019structure} & 1 \\
			\hline 
		\end{tabular}
	}
	\caption{\footnotesize Other predictor variables description table -- The \textit{id} column gives mnemonics according to data source, which is given in the second column \textit{Source}. The column \textit{frequency} states the sampling frequency of the variable. The column \textit{T-code} denotes the data transformation applied to a time-series, which are: (1) not transformed, (2) $\Delta x_t$, (3) $\Delta^2 x_t$, (4) log($x_t$), (5) $\Delta$ log ($x_t$), (6) $\Delta^2$ log ($x_t$). Panel A. describes real-time monthly macro series, panel B. describes daily financial markets data and panel C. monthly news attention series. \label{tab:other_data} }
\end{table}

\clearpage
{\scriptsize
	\begin{longtable}{r |ll cc}
		& Ticker & Firm name & PERMNO & RavenPack ID \\ 
		\hline
		1 & MMM & 3M & 22592 & 03B8CF \\ 
		2 & ABT & Abbott labs& 20482 & 520632 \\ 
		3 & AUD & Automatic data processing & 44644 & 66ECFD \\ 
		4 & ADTN & Adtran & 80791 & 9E98F2 \\ 
		5 & AEIS & Advanced energy industries & 82547 & 1D943E \\ 
		6 & AMG & Affiliated managers group & 85593 & 30E01D \\ 
		7 & AKST & A K steel holding & 80303 & 41588B \\ 
		8 & ATI & Allegheny technologies & 43123 & D1173F \\ 
		9 & AB & AllianceBernstein holding l.p. & 75278 & CB138D \\ 
		10 & ALL & Allstate corp. & 79323 & E1C16B \\ 
		11 & AMZN & Amazon.com & 84788 & 0157B1 \\ 
		12 & AMD & Advanced micro devices & 61241 & 69345C \\ 
		13 & DOX & Amdocs ltd. & 86144 & 45D153 \\ 
		14 & AMKR & Amkor technology & 86047 & 5C8D61 \\ 
		15 & APH & Amphenol corp. & 84769 & BB07E4 \\ 
		16 & AAPL & Apple & 14593 & D8442A \\ 
		17 & ADM & Archer daniels midland & 10516 & 2B7A40 \\ 
		18 & ARNC & Arconic & 24643 & EC821B \\ 
		19 & ATTA & AT\&T & 66093 & 251988 \\ 
		20 & AVY & Avery dennison corp. & 44601 & 662682 \\ 
		21 & BHI & Baker hughes & 75034 & 940C3D \\ 
		22 & BAC & Bank of america corp. & 59408 & 990AD0 \\ 
		23 & BAX & Baxter international inc. & 27887 & 1FAF22 \\ 
		24 & BBT & BB\&T corp. & 71563 & 1A3E1B \\ 
		25 & BDX & Becton dickinson \& co. & 39642 & 873DB9 \\ 
		26 & BBBY & Bed bath \& beyond inc. & 77659 & 9B71A7 \\ 
		27 & BHE & Benchmark electronics inc. & 76224 & 6CF43C \\ 
		28 & BA & Boeing co. & 19561 & 55438C \\ 
		29 & BK & Bank of new york mellon corp. & 49656 & EF5BED \\ 
		30 & BWA & BorgWarner inc. & 79545 & 1791E7 \\ 
		31 & BP & BP plc & 29890 & 2D469F \\ 
		32 & EAT & Brinker international inc. & 23297 & 732449 \\ 
		33 & BMY & Bristol-Myers squibb co. & 19393 & 94637C \\ 
		34 & BRKS & Brooks automation inc. & 81241 & FC01C0 \\ 
		35 & CA & CA technologies inc. & 25778 & 76DE40 \\ 
		36 & COG & Cabot oil \& gas corp. & 76082 & 388E00 \\ 
		37 & CDN & Cadence design systems inc. & 11403 & CC6FF5 \\ 
		38 & COF & Capital one financial corp. & 81055 & 055018 \\ 
		39 & CRR & Carbo ceramics inc. & 83366 & 8B66CE \\ 
		40 & CSL & Carlisle cos. & 27334 & 9548BB \\ 
		41 & CCL & Carnival corporation \& plc & 75154 & 067779 \\ 
		42 & CERN & Cerner corp. & 10909 & 9743E5 \\ 
		43 & CHRW & C.H. robinson worldwide inc. & 85459 & C659EB \\ 
		44 & SCHW & Charles schwab corp. & 75186 & D33D8C \\ 
		45 & CHKP & Check point software technologies ltd. & 83639 & 531EF1 \\ 
		46 & CHV & Chevron corp. & 14541 & D54E62 \\ 
		47 & CI & CIGNA corp. & 64186 & 86A1B9 \\ 
		48 & CTAS & Cintas corp. & 23660 & BFAEB4 \\ 
		49 & CLX & Clorox co. & 46578 & 719477 \\ 
		50 & KO & Coca-Cola co. & 11308 & EEA6B3 \\ 
		51 & CGNX & Cognex corp. & 75654 & 709AED \\ 
		52 & COLM & Columbia sportswear co. & 85863 & 5D0337 \\ 
		53 & CMA & Comerica inc. & 25081 & 8CF6DD \\ 
		54 & CRK & Comstock resources inc. & 11644 & 4D72C8 \\ 
		55 & CAG & ConAgra foods inc. & 56274 & FA40E2 \\ 
		56 & STZ & Constellation brands inc. & 69796 & 1D1B07 \\ 
		57 & CVG & Convergys corp. & 86305 & 914819 \\ 
		58 & COST & Costco wholesale corp. & 87055 & B8EF97 \\ 
		59 & CCI & Crown castle international corp. & 86339 & 275300 \\ 
		60 & DHR & Danaher corp. & 49680 & E124EB \\ 
		61 & DRI & Darden restaurants inc. & 81655 & 9BBFA5 \\ 
		62 & DVA & DaVita inc. & 82307 & EFD406 \\ 
		63 & DO & Diamond offshore drilling inc. & 82298 & 331BD2 \\ 
		64 & D & Dominion resources inc. & 64936 & 977A1E \\ 
		65 & DOV & Dover corp. & 25953 & 636639 \\ 
		66 & DOW & Dow chemical co. & 20626 & 523A06 \\ 
		67 & DHI & D.R. horton inc. & 77661 & 06EF42 \\ 
		68 & EMN & Eastman chemical co. & 80080 & D4070C \\ 
		69 & EBAY & eBay inc. & 86356 & 972356 \\ 
		70 & EOG & EOG resources inc. & 75825 & A43906 \\ 
		71 & EL & Estee lauder cos. inc. & 82642 & 14ED2B \\ 
		72 & ETH & Ethan allen interiors inc. & 79037 & 65CF8E \\ 
		73 & ETFC & E*TRADE financial corp. & 83862 & 28DEFA \\ 
		74 & XOM & Exxon mobil corp. & 11850 & E70531 \\ 
		75 & FII & Federated investors inc. & 86102 & 73C9E2 \\ 
		76 & FDX & FedEx corp. & 60628 & 6844D2 \\ 
		77 & FITB & Fifth third bancorp & 34746 & 8377DB \\ 
		78 & FISV & Fiserv inc. & 10696 & 190B91 \\ 
		79 & FLEX & Flex ltd. & 80329 & B4E00D \\ 
		80 & F & Ford motor co. & 25785 & A6213D \\ 
		81 & FWRD & Forward air corp. & 79841 & 10943B \\ 
		82 & BEN & Franklin resources inc. & 37584 & 5B6C11 \\ 
		83 & GE & General electric co. & 12060 & 1921DD \\ 
		84 & GIS & General mills inc. & 17144 & 9CA619 \\ 
		85 & GNTX & Gentex corp. & 38659 & CC339B \\ 
		86 & HAL & Halliburton Co. & 23819 & 2B49F4 \\ 
		87 & HLIT & Harmonic inc. & 81621 & DD9E41 \\ 
		88 & HIG & Hartford financial services group inc. & 82775 & 766047 \\ 
		89 & HAS & Hasbro inc. & 52978 & AA98ED \\ 
		90 & HLX & Helix energy solutions group inc. & 85168 & 6DD6BA \\ 
		91 & HP & Helmerich \& payne inc. & 32707 & 1DE526 \\ 
		92 & HSY & Hershey co. & 16600 & 9F03CF \\ 
		93 & HES & Hess corp. & 28484 & D0909F \\ 
		94 & HON & Honeywell international inc. & 10145 & FF6644 \\ 
		95 & JBHT & J.B. Hunt transport services Inc. & 42877 & 72DF04 \\ 
		96 & HBAN & Huntington bancshares inc. & 42906 & C9E107 \\ 
		97 & IBM & IBM corp. & 12490 & 8D4486 \\ 
		98 & IEX & IDEX corp. & 75591 & E8B21D \\ 
		99 & IR & Ingersoll-Rand plc & 12431 & 5A6336 \\ 
		100 & IDTI & Integrated device technology inc. & 44506 & 8A957F \\ 
		101 & INTC & Intel corp. & 59328 & 17EDA5 \\ 
		102 & IP & International paper co. & 21573 & 8E0E32 \\ 
		103 & IIN & ITT corp. & 12570 & 726EEA \\ 
		104 & JAKK & Jakks pacific inc. & 83520 & 5363A2 \\ 
		105 & JNJ & Johnson \& johnson & 22111 & A6828A \\ 
		106 & JPM & JPMorgan chase \& co. & 47896 & 619882 \\ 
		107 & K & Kellogg co. & 26825 & 9AF3DC \\ 
		108 & KMB & Kimberly-Clark corp. & 17750 & 3DE4D1 \\ 
		109 & KNGT & Knight transportation inc. & 80987 & ED9576 \\ 
		110 & LSTR & Landstar system inc. & 78981 & FD4E8D \\ 
		111 & LSCC & Lattice semiconductor corp. & 75854 & 8303CD \\ 
		112 & LLY & Eli lilly \& co. & 50876 & F30508 \\ 
		113 & LFUS & Littelfuse inc. & 77918 & D06755 \\ 
		114 & LNC & Lincoln national corp. & 49015 & 5C7601 \\ 
		115 & LMT & Lockheed martin corp. & 21178 & 96F126 \\ 
		116 & MTB & M\&T bank corp. & 35554 & D1AE3B \\ 
		117 & MANH & Manhattan associates inc. & 85992 & 031025 \\ 
		118 & MAN & ManpowerGroup inc. & 75285 & C0200F \\ 
		119 & MAR & Marriott international inc. & 85913 & 385DD4 \\ 
		120 & MMC & Marsh \& mcLennan cos. & 45751 & 9B5968 \\ 
		121 & MCD & McDonald's corp. & 43449 & 954E30 \\ 
		122 & MCK & McKesson corp. & 81061 & 4A5C8D \\ 
		123 & MDU & MDU resources group inc. & 23835 & 135B09 \\ 
		124 & MRK & Merck \& co. inc. & 22752 & 1EBF8D \\ 
		125 & MTOR & Meritor inc & 85349 & 00326E \\ 
		126 & MTG & MGIC investment corp. & 76804 & E28F22 \\ 
		127 & MGM & MGM resorts international & 11891 & 8E8E6E \\ 
		128 & MCHP & Microchip technology inc. & 78987 & CDFCC9 \\ 
		129 & MU & Micron technology inc. & 53613 & 49BBBC \\ 
		130 & MSFT & Microsoft corp. & 10107 & 228D42 \\ 
		131 & MOT & Motorola solutions inc. & 22779 & E49AA3 \\ 
		132 & MSM & MSC industrial direct co.& 82777 & 74E288 \\ 
		133 & MUR & Murphy oil corp. & 28345 & 949625 \\ 
		134 & NBR & Nabors industries ltd. & 29102 & E4E3B7 \\ 
		135 & NOI & National oilwell varco inc. & 84032 & 5D02B7 \\ 
		136 & NYT & New york times co. & 47466 & 875F41 \\ 
		137 & NFX & Newfield exploration co. & 79915 & 9C1A1F \\ 
		138 & NEM & Newmont mining corp. & 21207 & 911AB8 \\ 
		139 & NKE & NIKE inc. & 57665 & D64C6D \\ 
		140 & NBL & Noble energy inc. & 61815 & 704DAE \\ 
		141 & NOK & Nokia corp. & 87128 & C12ED9 \\ 
		142 & NOC & Northrop grumman corp. & 24766 & FC1B7B \\ 
		143 & NTRS & Northern trust corp. & 58246 & 3CCC90 \\ 
		144 & NUE & NuCor corp. & 34817 & 986AF6 \\ 
		145 & ODEP & Office depot inc. & 75573 & B66928 \\ 
		146 & ONB & Old national bancorp & 12068 & D8760C \\ 
		147 & OMC & Omnicom group inc. & 30681 & C8257F \\ 
		148 & OTEX & Open text corp. & 82833 & 34E891 \\ 
		149 & ORCL & Oracle corp. & 10104 & D6489C \\ 
		150 & ORBK & Orbotech ltd. & 78527 & 290820 \\ 
		151 & PCAR & Paccar inc. & 60506 & ACF77B \\ 
		152 & PRXL & Parexel international corp. & 82607 & EF8072 \\ 
		153 & PH & Parker hannifin corp. & 41355 & 6B5379 \\ 
		154 & PTEN & Patterson-uti energy inc. & 79857 & 57356F \\ 
		155 & PBCT & People's united financial inc. & 12073 & 449A26 \\ 
		156 & PEP & PepsiCo inc. & 13856 & 013528 \\ 
		157 & PFE & Pfizer inc. & 21936 & 267718 \\ 
		158 & PIR & Pier 1 imports inc. & 51692 & 170A6F \\ 
		159 & PXD & Pioneer natural resources co. & 75241 & 2920D5 \\ 
		160 & PNCF & PNC financial services group inc. & 60442 & 61B81B \\ 
		161 & POT & Potash corporation of saskatchewan inc. & 75844 & FFBF74 \\ 
		162 & PPG & PPG industries inc. & 22509 & 39FB23 \\ 
		163 & PX & Praxair inc. & 77768 & 285175 \\ 
		164 & PG & Procter \& gamble co. & 18163 & 2E61CC \\ 
		165 & PTC & PTC inc. & 75912 & D437C3 \\ 
		166 & PHM & PulteGroup inc. & 54148 & 7D5FD6 \\ 
		167 & QCOM & Qualcomm inc. & 77178 & CFF15D \\ 
		168 & DGX & Quest diagnostics inc. & 84373 & 5F9CE3 \\ 
		169 & RL & Ralph lauren corp. & 85072 & D69D42 \\ 
		170 & RTN & Raytheon co. & 24942 & 1981BF \\ 
		171 & RF & Regions financial corp. & 35044 & 73C521 \\ 
		172 & RCII & Rent-a-center inc. & 81222 & C4FBDC \\ 
		173 & RMD & ResMed inc. & 81736 & 434F38 \\ 
		174 & RHI & Robert half international inc. & 52230 & A4D173 \\ 
		175 & RDC & Rowan cos. inc. & 45495 & 3FFA00 \\ 
		176 & RCL & Royal caribbean cruises ltd. & 79145 & 751A74 \\ 
		177 & RPM & RPM international inc. & 65307 & F5D059 \\ 
		178 & RRD & RR R.R. donnelley \& sons co. & 38682 & 0BE0AE \\ 
		179 & SLB & Schlumberger ltd. n.v. & 14277 & 164D72 \\ 
		180 & SCTT & Scotts miracle-gro co. & 77300 & F3FCC3 \\ 
		181 & SM & SM st. mary land \& exploration co. & 78170 & 6A3C35 \\ 
		182 & SONC & Sonic corp. & 76568 & 80D368 \\ 
		183 & SO & Southern co. & 18411 & 147C38 \\ 
		184 & LUV & Southwest airlines co. & 58683 & E866D2 \\ 
		185 & SWK & Stanley black \& decker inc. & 43350 & CE1002 \\ 
		186 & STT & State street corp. & 72726 & 5BC2F4 \\ 
		187 & TGNA & TEGNA inc. & 47941 & D6EAA3 \\ 
		188 & TXN & Texas instruments inc. & 15579 & 39BFF6 \\ 
		189 & TMK & Torchmark corp. & 62308 & E90C84 \\ 
		190 & TRV & The travelers companies inc. & 59459 & E206B0 \\ 
		191 & TBI & TrueBlue inc. & 83671 & 9D5D35 \\ 
		192 & TUP & Tupperware brands corp. & 83462 & 2B0AF4 \\ 
		193 & TYC & Tyco international plc & 45356 & 99333F \\ 
		194 & TSN & Tyson foods inc. & 77730 & AD1ACF \\ 
		195 & X & United states Steel corp. & 76644 & 4E2D94 \\ 
		196 & UNH & UnitedHealth group inc. & 92655 & 205AD5 \\ 
		197 & VIAV & Viavi solutions inc. & 79879 & E592F0 \\ 
		198 & GWW & W.W. grainger inc. & 52695 & 6EB9DA \\ 
		199 & WDR & Waddell \& reed financial inc. & 85931 & 2F24A5 \\ 
		200 & WBA & Walgreens boots alliance inc. & 19502 & FACF19 \\ 
		201 & DIS & Walt disney co. & 26403 & A18D3C \\ 
		202 & WAT & Waters corp. & 82651 & 1F9D90 \\ 
		203 & WBS & Webster financial corp. & 10932 & B5766D \\ 
		204 & WFC & Wells fargo \& co. & 38703 & E8846E \\ 
		205 & WERN & Werner enterprises inc. & 10397 & D78BF1 \\ 
		206 & WABC & Westamerica bancorp & 82107 & 622037 \\ 
		207 & WDC & Western digital corp. & 66384 & CE96E7 \\ 
		208 & WHR & Whirlpool corp. & 25419 & BDD12C \\ 
		209 & WFM & Whole foods market inc. & 77281 & 319E7D \\ 
		210 & XLNX & Xilinx inc. & 76201 & 373E85 \\ 
		\hline
		\caption{\footnotesize Final list of firms -- The table contains the information about the full list of firms: tickers, firm names, CRSP PERMNO code and RavenPack ID. Tickers and firm names are taken as of June, 2017. PERMNO and RavenPack ID columns are used to match firms and firm news data. \label{tab:list_firms}} 
	\end{longtable}
}
\newpage

\section{Additional Empirical Results \label{appsec:emp}}

\begin{table}[h]
	\centering
	{\footnotesize
		\begin{tabular}{lccccc}
			& & & & & \\
			& $\hat pe_{i,t+1}$ & $\hat S$ &	& $\hat r_{i,t+1}$ & $ \hat e^a_{i,t+1|t}$\\
			& & & & & \\ 
			RW & 1.355  & & & 0.054 & 0.194 \\ 
			Consensus& 1.305  & & & & \\ 
			&\multicolumn{5}{c}{\textit{Individual}} \\ 
			BIC & 0.915 & 0.883 & &0.930& 0.819\\
			DM p-val RW  & 0.119&0.113 & & & \\
			AIC & 0.901 & 0.893 & &1.065 &0.875 \\ 
			DM p-val RW  & 0.118&0.115 & & & \\
			AICc & 0.926 & 0.902 & &1.073 & 0.896\\ 
			DM p-val RW  &0.123& 0.115 & & & \\
			&\multicolumn{5}{c}{\textit{Pooled}} \\  
			BIC & 0.894 & 0.790 & & 0.926 & 0.794 \\ 
			DM p-val RW  &0.060&0.026& & & \\
			AIC & 0.893 & 0.794 &  & 0.930 & 0.798\\ 
			DM p-val RW  & 0.058&0.027 & & & \\
			AICc & 0.893 & 0.795 & & 0.930 & 0.799 \\ 
			DM p-val RW  &0.058&0.028& & & \\
			&\multicolumn{5}{c}{\textit{Fixed effects}} \\
			BIC & 0.818 & 0.794 &  & 0.932 & 0.801  \\ 
			DM p-val RW  &0.053&0.032 & & & \\
			AIC & 0.814 & 0.797 & & 0.947 & 0.804 \\ 
			DM p-valat RW  &0.051 &0.034& & & \\
			AICc & 0.814 & 0.798 & & 0.947 & 0.804 \\
			DM p-val RW  &0.051&0.035 & & & \\ 
			\hline
	\end{tabular}}
	\caption{\footnotesize Column $\hat pe_{i,t+1}$ reports results for directly nowcasting the log P/E ratio, column $\hat S$ reports the results of nowcasting and summing up the components, column $r_{i,t+1}$ reports results for the log return and column $\hat e^a_{i,t+1|t}$ reports results for the log earnings forecast error of analysts. Row {\it RW} reports results for the random walk, while row {\it Consensus} for the median consensus nowcast. Panels {\it Individual},  {\it Pooled} and {\it Fixed effects} report results for different panel data models relative to the consensus MSE (columns $\hat pe_{i,t+1}$ and $\hat S$) and for the components (columns $r_{i,t+1}$ and $\hat e^a_{i,t+1|t}$) relative to the RW MSE. DM is the \cite{diebold1995comparing} test statistic p-values using one-sided critical values.  \label{apptab:nowcasts}
	}
\end{table}

\end{document}